 \documentclass[11pt]{amsart}
\usepackage{amsmath}

  \usepackage{paralist}
  \usepackage{graphics} 
  \usepackage{epsfig} 
\usepackage{graphicx}  \usepackage{epstopdf}
\usepackage{footnote}
\usepackage{timet}
\usepackage{algorithm}
\usepackage{algorithmic}
\usepackage{epsfig,subfigure}
\usepackage{epsf}
\usepackage{graphics}
\usepackage{url}
\usepackage{color}
\usepackage{amssymb}
\usepackage{amsmath}
\usepackage{float}
\usepackage{perpage}
\MakeSorted{figure}
\MakeSorted{table}
\usepackage{caption}

 \usepackage[colorlinks=true]{hyperref}
\hypersetup{urlcolor=blue, citecolor=red}

\addtocounter{footnote}{0}
\newcommand{\bfy}{\mathbf{y}}
\newcommand{\bfb}{\mathbf{b}}
\newcommand{\bfu}{\mathbf{u}}
\newcommand{\bfv}{\mathbf{v}}
\newcommand{\bfA}{\mathbf{A}}
\newcommand{\bfB}{\mathbf{B}}
\newcommand{\bfC}{\mathbf{C}}
\newcommand{\bfD}{\mathbf{D}}
\newcommand{\bfH}{\mathbf{H}}
\newcommand{\bfJ}{\mathbf{J}}

\newcommand{\bfM}{\mathbf{M}}
\newcommand{\bfQ}{\mathbf{Q}}

\newcommand{\bfT}{\mathbf{T}}
\newcommand{\bfU}{\mathbf{U}}
\newcommand{\bfV}{\mathbf{V}}
\newcommand{\bfW}{\mathbf{W}}
\newcommand{\bfX}{\mathbf{X}}
\newcommand{\bfY}{\mathbf{Y}}
\newcommand{\bfSigma}{\mathbf{\Sigma}}
\newcommand{\bfTj}{\mathbf{T}_j}
\newcommand{\bfXD}{\mathbf{X}_{D}}
\newcommand{\bfXS}{\mathbf{X}_{S}}
\newcommand{\bfTD}{\mathbf{T}_{D}}
\newcommand{\bfTS}{\mathbf{T}_{S}}

\newcommand{\rank}{\mathtt{rank}}
\newcommand{\fft}{\mathtt{fft}}
\newcommand{\ifft}{\mathtt{ifft}}
\newcommand{\myvec}{\mathtt{vec}}

\newcommand{\extract}{\mathtt{extract}}
\newcommand{\subjectto}{\mathtt{subject\:to}}
\newcommand{\diag}{\mathtt{diag}}
\newcommand{\xmn}{x_{mn}}
\newcommand{\xonej}{x_{1j}}
\newcommand{\xtwoj}{x_{2j}}
\newcommand{\xthreej}{x_{3j}}
\newcommand{\xKj}{x_{Kj}}
\newcommand{\xKplusonej}{x_{(K+1)j}}
\newcommand{\RPQ}{\mathcal{R}^{P \times Q}}
\newcommand{\RPcQc}{\mathcal{R}^{P_{C} \times Q_{C}}}

\title[FNCLRMD\_PFS] 
      {Fast non-convex low-rank matrix decomposition for separation of potential field data using minimal memory}

\author[Dan Zhu, Rosemary A. Renaut, Hongwei Li and Tianyou Liu]{}


 \email{zhud\_igg@cug.edu.cn}
 \email{renaut@asu.edu}
 \email{hwli@cug.edu.cn}
 \email{liuty@cug.edu.cn}

\thanks{The first author is supported by the National Key R \& D Program of China 2018YFC1503705; The second author is supported by the NSF grant DMS 1913136; The third author is supported by the National Key R \& D Program of China 2018YFC1503705 and Hubei Subsurface Multi-scale Imaging Key Laboratory (China University of Geosciences) SMIL-2018-06;}

\begin{document}
\maketitle

\centerline{\scshape Dan Zhu}
\medskip
{\footnotesize
 \centerline{Institude of Geophysics \& Geomatics}
   \centerline{China University of Geosciences (Wuhan)}
   \centerline{ Wuhan, MO 430074, China }
} 

\medskip

\centerline{\scshape Rosemary A. Renaut$^*$}
\medskip
{\footnotesize
 \centerline{ School of Mathematical and Statistical Sciences}
   \centerline{Arizona State University}
   \centerline{Tempe, MO 85287, USA}
}

\medskip

\centerline{\scshape Hongwei Li and Tianyou Liu}
\medskip
{\footnotesize
 \centerline{Institude of Geophysics \& Geomatics}
   \centerline{China University of Geosciences (Wuhan)}
   \centerline{ Wuhan, MO 430074, China }

\bigskip


\begin{abstract}
A fast non-convex low-rank matrix decomposition method for potential field data separation is proposed. The singular value decomposition of the large size trajectory matrix, which is also a block Hankel matrix, is obtained using a fast randomized singular value decomposition algorithm in which  fast block Hankel matrix-vector multiplications are implemented with minimal memory storage. This  fast block Hankel matrix randomized singular value decomposition algorithm is integrated into the \texttt{Altproj} algorithm, which is a standard non-convex method for solving the robust principal component analysis optimization problem. The improved algorithm avoids the construction of the trajectory matrix. Hence, gravity and magnetic data matrices of large size can be computed. Moreover, it is more efficient than   the traditional low-rank matrix decomposition method, which is based on the use of an inexact augmented Lagrange multiplier algorithm. The presented algorithm is also  robust and, hence, algorithm-dependent parameters are easily determined. The improved and traditional algorithms are contrasted for the separation of synthetic gravity and magnetic data matrices of different sizes. The presented results demonstrate that the improved algorithm is not only computationally more efficient but it is also more accurate. Moreover, it is possible to solve far larger problems. As an example, for the adopted computational environment, matrices of sizes larger than $205 \times 205$ generate ``out of memory" exceptions with the traditional method, but a matrix of size $2001\times 2001$ can be calculated in   $1062.29$s with the new algorithm.   Finally, the improved method is applied to separate real gravity and magnetic data in the Tongling area, Anhui province, China. Areas which may exhibit mineralizations are inferred based on the separated anomalies.
\end{abstract}


\section{Introduction}
\label{sec:intro}

To study target geological sources, the target gravity, or magnetic anomalies, that are caused by the target sources, should be separated from the total fields which are the superposition of the gravity and magnetic fields caused by all underground sources.  Separated anomalies are then used for data inversion and interpretation of  geological features. Therefore, the separation of potential field data is an important step for high quality inversion and interpretation. Deep sources generate large scale smooth anomalies which are called regional anomalies. Residual anomalies, which are on a small scale, are caused by shallow sources. There are many methods for separating the regional-residual anomalies. They can be classified into three types. The classical methods of the first group separate the data in the spatial domain. These include methods such as the moving average, polynomial fitting, minimum curvature, and empirical mode decomposition,  \cite{W.M.Telford2003,Agocs1951,Mickus1991,Mandal2018}. Methods of the second and third types separate the anomalies in the frequency or wavelet domains, respectively. These include methods such as matched filtering, Wiener filtering, continuation, and discrete wavelet analysis, \cite{Clarke1969,Pawlowski1995,Spector1970,Pawlowski1990,Fedi1998}. While algorithms that  separate the anomalies in the frequency or wavelet domains are  easy to  implement \cite{Hou1997Wavelet,Yang2001Discrete}, the spectral overlapping of the regional and residual anomalies makes it difficult to obtain satisfactory results \cite{zhang2009separation}.

It has been demonstrated in areas of image and signals processing  that  the use of a low-rank matrix decomposition for  robust principal component analysis (\texttt{RPCA})   is very effective \cite{candes2011robust}. The fundamental observation is that practical data from applied science fields  is usually distributed on low-dimensional manifolds in high-dimensional spaces \cite{Lin2017}. The mathematical model for \texttt{RPCA} is a double-objective optimization that separates the matrix into a low-rank matrix and a sparse matrix. Because   \texttt{RPCA} is robust and provides high accuracy separation, it has been applied in many fields, and there is much research on solving the optimization problem. Generally, the Lagrange function is used to transform the double-objective optimization problem into a single-objective optimization problem that is solved using convex optimization. Iterative thresholding, accelerated proximal gradient, exact augmented Lagrange multiplier (\texttt{EALM}), and inexact augmented Lagrange multiplier (\texttt{IALM}) algorithms have been proposed to solve the convex optimization problem \cite{wright2010sparse,beck2009fast,lin2010augmented}. Due to the high computational cost of convex \texttt{RPCA}, a non-convex \texttt{RPCA} algorithm, which is called \texttt{Altproj}, has been proposed to reduce the cost \cite{netrapalli2014non}. Furthermore, as compared with convex \texttt{RPCA}, the higher accuracy of  \texttt{Altproj} has lead to its wide adoption.

A low-rank matrix decomposition algorithm for potential field separation (\texttt{LRMD\_PFS}),   based on \texttt{RPCA} and singular spectrum analysis,  has been proposed \cite{zhu2019low}. Singular spectrum analysis is a classical method using the trajectory matrix and the singular value decomposition (\texttt{SVD})  \cite{takens1981detecting,broomhead1986extracting,tsonis1996mapping}. An important step in  \texttt{LRMD\_PFS} is the construction of the trajectory matrix (which is a block Hankel matrix) of the total field. Then, the trajectory matrix of the total field can be separated into a low-rank matrix and a sparse matrix using convex \texttt{RPCA}. The separated low-rank   and  sparse matrices are the approximations of the trajectory matrices of the regional anomalies and the residual anomalies, respectively. The sparse features of the regional anomalies in the frequency domain, and the localization features of the residual anomalies in the spatial domain, are both considered in  \texttt{LRMD\_PFS}. Although \texttt{LRMD\_PFS} separates the anomalies without the use of a Fourier transform to the frequency domain, it can also be seen as providing a new group of methods because it provides a combination of the features of the potential field data in both spatial and frequency domains. Hence, as compared to classical methods, \texttt{LRMD\_PFS} is more robust and has higher accuracy. The computational cost of  \texttt{LRMD\_PFS} is, however, high. There is a large memory demand associated with generating and storing the large scale trajectory matrix, and a large number of operations are required to generate the \texttt{SVD} of a large matrix. For example, if the size of the matrix is $101\times101$, then the size of the constructed trajectory matrix is $2601\times2601$. The trajectory matrix then requires memory that is $663$ times that of the original data. For a matrix of size $201\times201$, the size of the trajectory matrix is $10201\times10201$, and the memory demand increases by a factor of almost $2576$. Therefore, the size of the trajectory matrix increases rapidly with the size of the original matrix.

In this paper, a fast block Hankel matrix randomized \texttt{SVD} (\texttt{FBHMRSVD}) algorithm  that requires minimal memory storage  is proposed.  \texttt{FBHMRSVD} is based on fast block Hankel matrix-vector multiplications (\texttt{FBHMVM}) \cite{Vogel2002,lu2015fast} and the use of a randomized \texttt{SVD} (\texttt{RSVD}) \cite{liberty2007randomized,halko2011finding,vatankhah2018fast}. This then yields a fast non-convex low-rank matrix decomposition for potential field separation (\texttt{FNCLRMD\_PFS}) that is based on the \texttt{FBHMRSVD}.  \texttt{FBHMRSVD} is used to approximate the \texttt{SVD} of the trajectory matrix without constructing the large trajectory matrix. Further, implementing  \texttt{FBHMRSVD} within \texttt{Altproj} also yields approximation of the trajectory matrices of the regional anomalies and residual anomalies without explicit construction of the trajectory matrix. Therefore,  the large scale potential field data matrix can be separated using \texttt{FNCLRMD\_PFS}. Furthermore,  \texttt{FNCLRMD\_PFS} has lower computational cost and higher accuracy than   \texttt{LRMD\_PFS}. The algorithm is developed and then contrasted with the classical approach for separation of synthetic data sets in Sections~\ref{FBHMRSVDSEC}-\ref{SyntheticResults}. Results showing that the algorithm efficiently and effectively separates real gravity and magnetic data in the Tongling area, Anhui province, China are presented in Section~\ref{PracticalData}.

\section{The Fast block Hankel matrix randomized singular value decomposition: \texttt{FBHMRSVD}}\label{FBHMRSVDSEC}
\subsection{Fast block Hankel matrix-vector multiplication: \texttt{FBHMVM}}

Consider a $2$D gridded potential field data matrix $\bfX=[\xmn] \in \RPQ$, where $\xmn$ denotes the element at the $m$th row and $n$th column of the matrix $\bfX$. Before constructing the trajectory matrix, the Hankel matrix $\bfTj$ is constructed using the $j$th column of $\bfX$ as follows,
\bigskip
\begin{eqnarray*}
\bfTj=\begin{bmatrix}
\xonej & \xtwoj & \dotsm & x_{(P-K+1)j}\\ 
\xtwoj & \xthreej & \dotsm & x_{(P-K+2)j}\\ 
\vdots & \vdots &  & \vdots\\ 
\xKj & \xKplusonej & \dotsm & x_{Pj}
\end{bmatrix}.
\end{eqnarray*}
\bigskip
Here, generally,  $K=\lfloor(P+1)/2\rfloor$, where $\lfloor\:\rfloor$ denotes the integer part of its argument. If  $\bfTj$ has size $K \times L$, then $L=P-K+1$, and trajectory matrix $\bfT$ of size $K\hat{K} \times L (Q-\hat{K}+1)$ is constructed as follows,
\bigskip
\begin{eqnarray*}
\bfT=\begin{bmatrix}
\bfT_1 & \bfT_2 & \dotsm & \bfT_{Q-\hat{K}+1}\\ 
\bfT_2 & \bfT_3 & \dotsm & \bfT_{Q-\hat{K}+2}\\ 
\vdots & \vdots &  & \vdots\\ 
\bfT_{\hat{K}} & \bfT_{\hat{K}+1} & \dotsm & \bfT_{Q}
\end{bmatrix}.
\end{eqnarray*}
\bigskip
Setting $\hat{K} = \lfloor (Q+1)/2 \rfloor$ makes $\bfT$ as near to square as possible. $\bfT$ is a block Hankel matrix    with $\hat{K} \times \hat{L}$ blocks, where $\hat{L} = Q - \hat{K} + 1$. The construction of $\bfT$ from $\bfX$ is denoted by 
\bigskip
\begin{eqnarray}\label{process of T construction}
\bfT=\mathcal{H}(\bfX).
\end{eqnarray}
\bigskip

Now, given a block Hankel matrix, efficient evaluation of  matrix-vector products
\bigskip
\begin{eqnarray}\label{y=Tb}
\bfy=\bfT\bfb,
\end{eqnarray}
\bigskip
is required. Direct evaluation of the matrix-vector product  using first  \eqref{process of T construction} to find $\bfT$ and then calculating \eqref{y=Tb}, uses approximately $8K \hat{K} L \hat{L} - K \hat{K} - L \hat{L}$ flops  and requires storage of $K \hat{K} L \hat{L} + K \hat{K} + L \hat{L}$ floating point entries.  On the other hand, using Algorithm~\ref{FBHMVMAlgorithm}, $\bfy$ can be calculated from $\bfX$ and   $\bfb$ without constructing   $\bfT$  using  $\mathcal{O}(PQ\log_{2}{PQ})$ flops and a storage requirement of $3PQ+K \hat{K} + L \hat{L}$ entries. Here, the fast operation that combines \eqref{process of T construction} and \eqref{y=Tb} is detailed in Algorithm~\ref{FBHMVMAlgorithm}, and  is denoted by
\bigskip
\begin{eqnarray*}
\bfy=\mathtt{FBHMVM}(\bfX,\bfb,K,\hat{K}).
\end{eqnarray*}

\begin{algorithm}
\caption{Fast block Hankel matrix-vector multiplication: $\bfy=\mathtt{FBHMVM}(\bfX,\bfb,K,\hat{K})$.}\label{FBHMVMAlgorithm}
\begin{algorithmic}[1]
\STATE \label{VMstep1}\textbf{Input:} potential field data matrix $\bfX \in \RPQ$; vector $\bfb$.
\STATE $\hat{\bfW}=\ifft2(\fft2(\bfT^{circ}).*\fft2 (\bfW))$.
\STATE \textbf{Output:} $\bfy=\bfJ\myvec(\extract(\hat{\bfW}))$.
\end{algorithmic}
\end{algorithm}

Algorithm~\ref{FBHMVMAlgorithm} uses the exchange matrix $\bfJ$. This is the permutation matrix which is $0$ everywhere except for $1$s on the counter diagonal. It is also referred to as the reversal matrix, backward identity, or standard involutory permutation matrix. $\bfT^{circ}$ is defined by
\bigskip
\begin{eqnarray*}
\bfT^{circ}=\begin{bmatrix}
\hat{\bfT}_{\hat{K}}& \dotsm & \hat{\bfT}_{1} & \hat{\bfT}_{\hat{L}} & \dotsm & \hat{\bfT}_{2}
\end{bmatrix},
\end{eqnarray*}
\bigskip
where $\hat{\bfT}_{j}$ is embedded from the $j$th column of $\bfX$ as follows,
\bigskip
\begin{eqnarray*}
\hat{\bfT}_{j}=
\begin{bmatrix}
x_{Kj}& \dotsm & x_{1j} & x_{Lj} & \dotsm & x_{2j}
\end{bmatrix}^{T}.
\end{eqnarray*}
\bigskip
$\bfW$ is constructed from $\bfb$ as follows,
\bigskip
\begin{eqnarray*}
\bfW=
\begin{bmatrix}
\bfB & \mathbf{0}_{L(\hat{L}-1)}\ \\
\mathbf{0}_{(L-1)\hat{L}} & \mathbf{0}_{(L-1)(\hat{L}-1)}
\end{bmatrix}^{T}.
\end{eqnarray*}
\bigskip
where $\bfb=\myvec(\bfB)$, and the operation $\myvec(\cdot)$ denotes the vectorization operation. Moreover, the extraction operation is defined by
\bigskip
\begin{eqnarray*}
\extract(\hat{\bfW})=\hat{\bfW}(1:L,1:\hat{L}).
\end{eqnarray*}

\subsection{Fast block Hankel matrix-matrix multiplication: \texttt{FBHMMM}}
It is immediate that Algorithm~\ref{FBHMVMAlgorithm} can be extended for block Hankel matrix-matrix multiplication 
\bigskip
\begin{eqnarray*}
\bfY=\bfT\bfC,
\end{eqnarray*}
\bigskip
where the size of $\bfC$ is $P_{C}\times Q_{C}$. The process is given in Algorithm~\ref{FBHMMMAlgorithm}, and  is denoted by
\bigskip
\begin{eqnarray*}
\bfY=\mathtt{FBHMMM}(\bfX,\bfC,K,\hat{K}).
\end{eqnarray*}

\begin{algorithm}
\caption{Fast block Hankel matrix-matrix multiplication: $\bfY=\mathtt{FBHMMM}(\bfX,\bfC,K,\hat{K})$.}\label{FBHMMMAlgorithm}
\begin{algorithmic}[1]
\STATE \textbf{Input:} potential field data matrix $\bfX \in \RPQ$; matrix $\bfC \in \RPcQc$; parameters $K$ and $\hat{K}$.
\FOR {$j=1:Q_{C}$.} 
\STATE $\bfY(:,j)=\mathtt{FBHMVM}(\bfX,\bfC(:,j),K,\hat{K})$.
\ENDFOR
\STATE \textbf{Output:} $\bfY$.
\end{algorithmic}
\end{algorithm}

\subsection{The fast block Hankel matrix randomized SVD: \texttt{FBHMRSVD} }
The \texttt{SVD} is the basis of matrix rank reduction, and it is an important step in \texttt{RPCA}. The process of the \texttt{SVD} for the block Hankel matrix is represented by
\bigskip
\begin{eqnarray*}
[\bfU,\bfSigma,\bfV]=\mathtt{SVD}(\mathcal{H}(\bfX)).
\end{eqnarray*}
\bigskip
Here $\bfU=[\bfu_{1},\bfu_{2},\dotsm]$ and $\bfV=[\bfv_{1},\bfv_{2},\dotsm]$ are unitary matrices, $\bfu_{1},\bfu_{2},\dotsm$ are left singular vectors, $\bfv_{1},\bfv_{2},\dotsm$ are right singular vectors; and $\bfSigma=\diag(\sigma_{1}^{2},\sigma_{2}^{2},\dotsm)$ is a diagonal matrix where $\sigma_1\geqslant \sigma_2 \geqslant \dotsm \geqslant 0$ are the singular values of $\bfT$, and
\bigskip
\begin{eqnarray*}
\bfT=\bfU\bfSigma\bfV^{T}.
\end{eqnarray*}
\bigskip
The cost of obtaining all terms of the \texttt{SVD} is $\mathcal{O}((L\hat{L})(K\hat{K})^2)$, \cite{GoLo:96}, which can be prohibitive when $P$ and $Q$ are large. For the rank reduction problem, however, not all terms are required and it can be sufficient to obtain a partial \texttt{SVD} with $r$ terms, corresponding to using a rank $r$ approximation,
\bigskip
\begin{eqnarray}\label{T=USigmaV_{T}}
\bfT_r= \bfU_r\bfSigma_r\bfV_r^{T},
\end{eqnarray}
\bigskip
where $\rank(\bfT_r)=r$. Generally,  low-rank features of $\bfT$ are required and so $r$ is relatively small.  Still, the cost of finding the exact dominant $r$ terms in  \eqref{T=USigmaV_{T}} is high. On the other hand,  the randomized singular value decomposition (\texttt{RSVD}), \cite{liberty2007randomized,halko2011finding}, has been proposed for efficient determination of a low rank matrix approximation $\bfT_r$ without the exact calculation of the components in \eqref{T=USigmaV_{T}}. Here, we implement the \texttt{RSVD} by taking advantage of all steps employing matrix-matrix multiplications with $\bfT$ using Algorithm~\ref{FBHMMMAlgorithm}, and  without explicitly obtaining $\bfT$.  This process, given in Algorithm~\ref{FBHMRSVDAlgorithm},   is denoted by
\bigskip
\begin{eqnarray*}
[\bfU_{r},\bfSigma_{r},\bfV_{r}]=\mathtt{FBHMRSVD}(\bfX,K,\hat{K},r,p,q).
\end{eqnarray*}
\bigskip

The integer parameters $p$, and $q$ are integral to the implementation of an \texttt{RSVD} algorithm. They represent an oversampling and  power iteration parameter, respectively.  When the required rank $r$ is relatively small with respect to the full rank of the matrix, it is sufficient to take $p=r$. While the accuracy of \texttt{RSVD} increases with increasing $q$,  the cost also increases. But if the spectrum separates into a dominant larger set of values with $\sigma_\ell \gg \sigma_{\ell+1}$, it is sufficient to use a relatively small $q$, such as $q=0$, $1$ or $2$, where $q>0$ applies $q$ steps of  a power iteration to improve the approximation to the dominant singular values. 

\begin{algorithm}
\caption{Fast block Hankel matrix \texttt{RSVD}: $[\bfU_{r},\bfSigma_{r},\bfV_{r}]=\mathtt{FBHMRSVD}(\bfX,K,\hat{K},r,p,q)$.}\label{FBHMRSVDAlgorithm}
\begin{algorithmic}[1]
\STATE \textbf{Input:} potential field data matrix $\bfX \in \RPQ$; desired rank $r$; oversampling parameter $p$; power iteration parameter $q$; parameter $K$ and $\hat{K}$.
\STATE $\ell=r+p$, $k=0$.
\STATE Generate a Gaussian random matrix $\mathbf{\Omega} \in \mathcal{R}^{\ell \times K\hat{K}}$.
\STATE \label{step4} $\bfA^{(0)}=\mathtt{FBHMMM}(\bfX,\mathbf{\Omega}^{T},L,\hat{L})$.
\STATE \label{step5}QR factorization: $[\bfQ^{(0)},\sim]=\mathtt{qr}(\bfA^{(0)})$, where $\bfQ^{(0)} \in \mathcal{R}^{L\hat{L} \times \ell}$ is an orthonormal matrix.
\WHILE {$q>k$}
\STATE \label{step7}$\bfA^{(1)}=\mathtt{FBHMMM}(\bfX,\mathbf{\bfQ}^{(0)},K,\hat{K})$.
\STATE \label{step8}$[\bfQ^{(1)},\sim]=\mathtt{qr}(\bfA^{(1)})$.
\STATE \label{step9}$\bfA^{(2)}=\mathtt{FBHMMM}(\bfX,\mathbf{\bfQ}^{(1)},L,\hat{L})$.
\STATE \label{step10}$[\bfQ^{(2)},\sim]=\mathtt{qr}(\bfA^{(2)})$.
\STATE $\bfQ^{(0)}=\bfQ^{(2)},k=k+1$.
\ENDWHILE 
\STATE \label{step13}$\bfB=\mathtt{FBHMMM}(\bfX,\mathbf{\bfQ}^{(0)},K,\hat{K})$.
\STATE \label{step14}Compute the eigen-decomposition of $\bfB^{T}\bfB$: $[\tilde{\bfV}_\ell,\bfD_\ell]=\mathtt{eig}(\bfB^{T}\bfB)$. 
\STATE \label{step15}$\bfV_r=\bfQ^{(0)}\tilde{\bfV}_\ell(:,1:r)$; $\mathbf{\sigma}_r=\sqrt{\bfD_\ell(1:r,1:r)}$; $\bfU_r=\bfB\tilde{\bfV}_\ell(:,1:r)\mathbf{\Sigma}_r^{-1}$.
\STATE \textbf{Output:} $\bfU_r$, $\mathbf{\Sigma}_r$, $\bfV_r$.
\end{algorithmic}
\end{algorithm}

In Algorithm~\ref{FBHMRSVDAlgorithm},  note that  steps \ref{step4}, \ref{step7}, \ref{step9}, and \ref{step13} involve  trajectory matrix-matrix multiplications and are replaced by the use of Algorithm~\ref{FBHMMMAlgorithm} in order to avoid calculation of $\bfT$. The original equations are
\bigskip
\begin{eqnarray*}
\bfA^{(0)}=\bfT^T\mathbf{\Omega}^T,\quad
\bfA^{(1)}=\bfT\bfQ^{(0)}, \quad
\bfA^{(2)}= \bfT^T\bfQ^{(1)} \text{ and   } 
\bfB=\bfT\bfQ^{(0)},
\end{eqnarray*}
\bigskip
where $\bfA^{(0)}$, $\bfA^{(2)} \in \mathcal{R}^{L\hat{L}\times \ell}$ and $\bfA^{(1)}$, $\bfQ^{(1)}$, $\bfB \in \mathcal{R}^{K\hat{K}\times \ell}$. 
Because Algorithm~\ref{FBHMRSVDAlgorithm} can be recast without using Algorithm~\ref{FBHMMMAlgorithm} for matrix-matrix multiplications, the accuracy of the two algorithms is the same, up to floating point arithmetic errors that may accrue. But the computational cost is much reduced. The computational cost in terms of flops and storage for each algorithm are detailed, for each step,   in Table~\ref{tab1}. The  storage and flops required for steps \ref{step5}, \ref{step8}, \ref{step10},  \ref{step14}, and \ref{step15} are the same.  For $K\hat{K}L\hat{L}\gg P Q$, however, these costs are far lower in steps \ref{step4}, \ref{step7}, \ref{step9}, and \ref{step13} when implemented using Algorithm~\ref{FBHMMMAlgorithm}.

\begin{table}
\caption{The computational cost measured in terms of floating point operations and storage of floating point entries at each step of Algorithm~\ref{FBHMRSVDAlgorithm} implemented with (\texttt{FBHMRSVD}), and without (\texttt{RSVD}), the use of \texttt{FBHMMM} for multiplications with $\bfT$.}\label{tab1}
\begin{tabular}{c | c c | c c}
\hline
\multicolumn{1}{c}{}&\multicolumn{2}{c}{\texttt{FBHMRSVD}}&\multicolumn{2}{c}{\texttt{RSVD}}\\ \hline
Step & Cost in flops&Cost in storage & Cost in flops&Cost in storage\\  \hline
\ref{step4}, \ref{step9}& $\mathcal{O}(\ell PQ\log_{2}{PQ})$&$PQ+\ell K\hat{K}+\ell L\hat{L}$& $2\ell K\hat{K}L\hat{L}$&$K\hat{K}L\hat{L}+\ell K\hat{K}+\ell L\hat{L}$\\
\ref{step5}, \ref{step10}& $2\ell^2(L\hat{L}-\ell /3)$&$2\ell L\hat{L}$& $2\ell^2(L\hat{L}-\ell /3)$&$2\ell L\hat{L}$\\
\ref{step7}, \ref{step13}& $\mathcal{O}(\ell PQ\log_{2}{PQ})$&$PQ+\ell K\hat{K}+\ell L\hat{L}$& $4\ell K\hat{K}L\hat{L}$&$K\hat{K}L\hat{L}+\ell K\hat{K}+\ell L\hat{L}$\\
\ref{step8}& $2\ell^2(K\hat{K}-\ell /3)$&$2\ell K\hat{K}$& $2\ell^2(K\hat{K}-\ell /3)$&$2\ell K\hat{K}$\\
\ref{step14}& $\mathcal{O}(\ell^3)$&$2\ell^2+\ell $& $\mathcal{O}(\ell^3)$&$2\ell^2+\ell $\\
\ref{step15}& $\ell r(2\ell +3K\hat{K})$&$r(K\hat{K}+L\hat{L})+2\ell^2$& $\ell r(2\ell +3K\hat{K})$&$r(K\hat{K}+L\hat{L})+2\ell^2$\\
 & &$+\ell L\hat{L}+r+\ell $& &$+\ell L\hat{L}+r+\ell $\\
\hline
\end{tabular}
\end{table}

\subsection{Experiments on \texttt{FBHMRSVD}}
We now discuss the influence of the parameters on the accuracy and computational costs of Algorithm~\ref{FBHMRSVDAlgorithm}. We compare the computational costs with, and without, the use of Algorithm~\ref{FBHMMMAlgorithm} for matrix multiplication, and the accuracy as compared to the use of the partial \texttt{SVD}. Hence, computations reported using \texttt{RSVD} and partial \texttt{SVD} are all based on the constructions of the trajectory matrices. The CPU of the computer for the computations in this paper is the Intel(R) Xeon (R) Gold 6138 CPU @ 2.00GHz; the release of MATLAB is 2019b.

First, we discuss the influence of the parameters on the computational cost and accuracy of Algorithm~\ref{FBHMRSVDAlgorithm}. Figure~\ref{fig1} shows the influence of the parameter $q$ on different sizes of the matrix $\bfX$, for $r=10$ and $q=0$, $1$ and $2$. The improvement in reducing the root mean square error (\texttt{RMSE}), which is defined by $\|\bfX^*-\hat{\bfX}\|_2/PQ$ where $\bfX^*$ and $\hat{\bfX}$ are the rank-$r$ approximations of $\bfX$ using the full SVD and Algorithm~\ref{FBHMRSVDAlgorithm}, respectively, is most significant when one power iteration is introduced, $q=1$.  With larger $q$ the computational cost increases, as can be seen from Figure~\ref{fig1}, demonstrating that the cost more than doubles when going from $q=0$ to $q=1$, but does not quite double again going from $q=1$ to $q=2$. Clearly there is a trade off between cost and accuracy, thus we recommend $q=1$ as a suitable compromise.

\begin{figure}
\subfigure[]{\label{fig1a}\includegraphics[width=.65\textwidth]{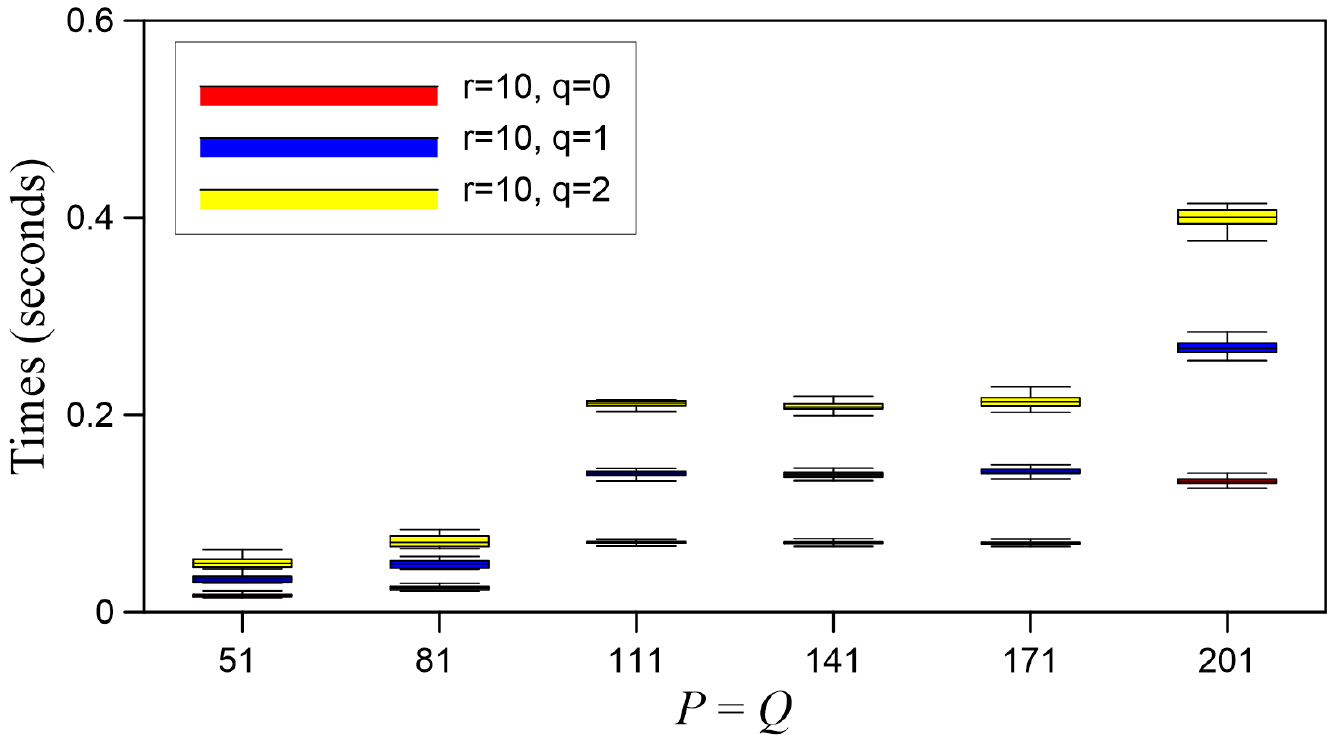}}\\
\subfigure[]{\label{fig1b}\includegraphics[width=.65\textwidth]{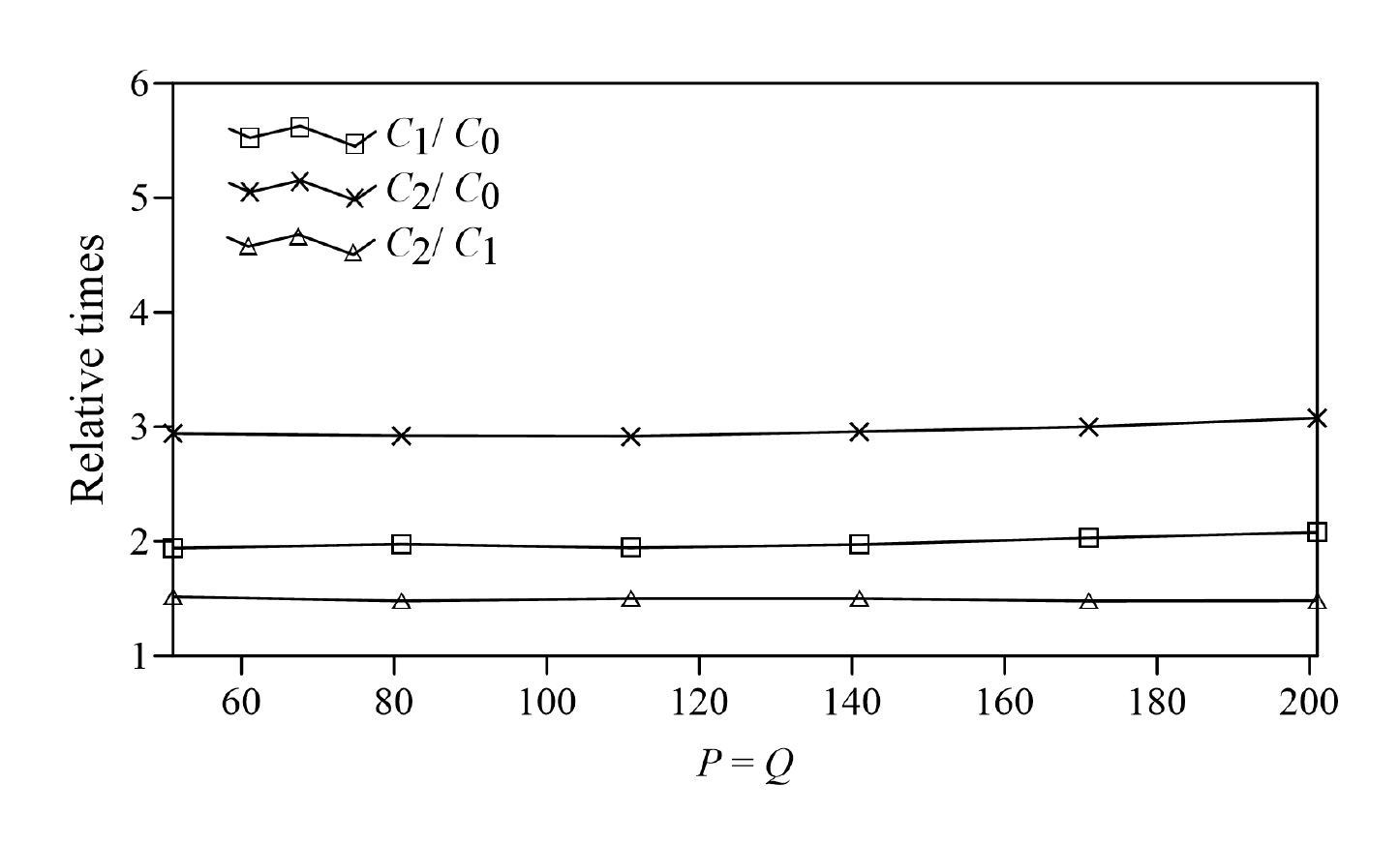}}\\
\subfigure[]{\label{fig1c}\includegraphics[width=.65\textwidth]{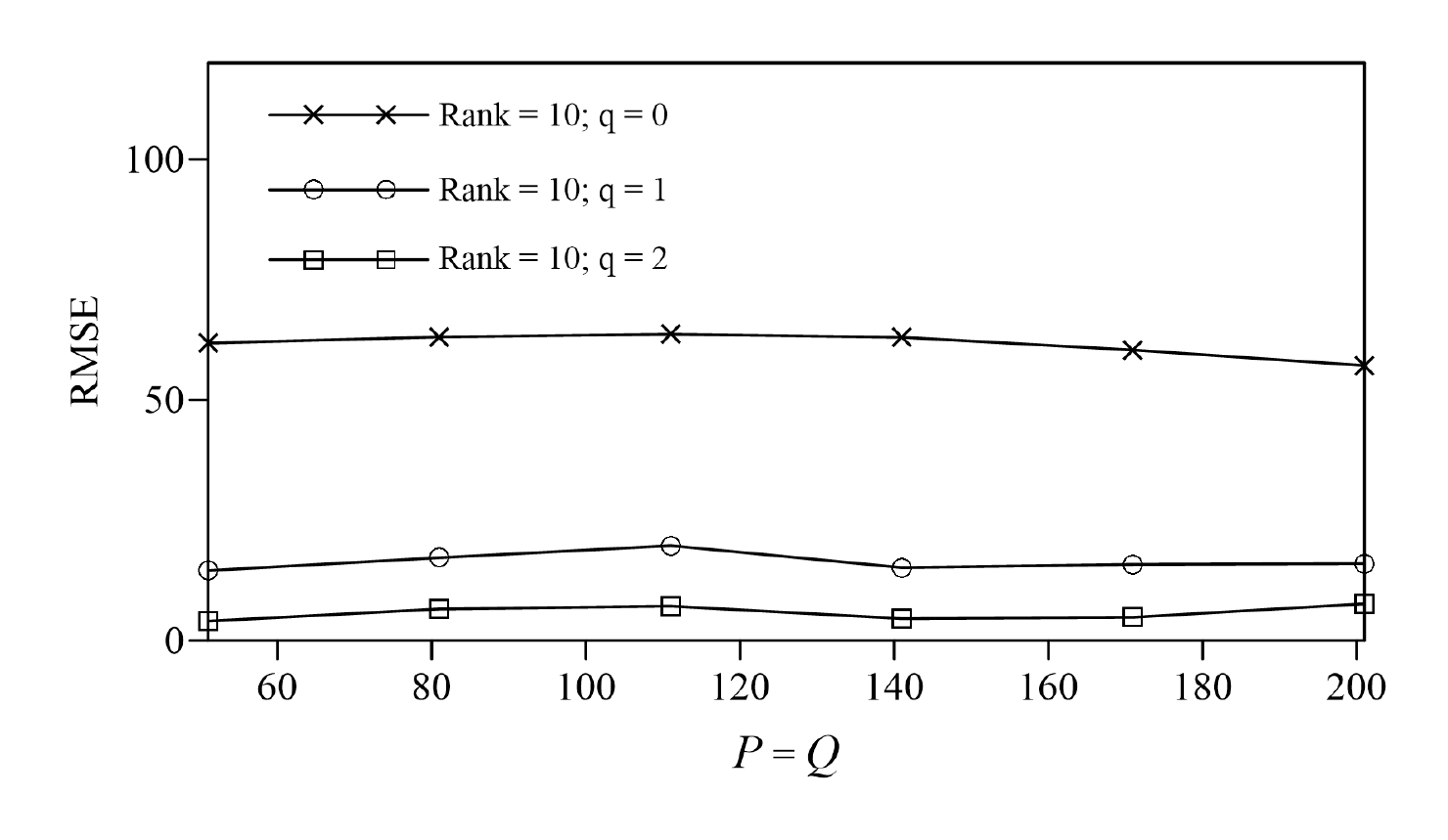}}\\
\caption{Experiments for Algorithm~\ref{FBHMRSVDAlgorithm} for $r=10$ with $q=0$, $1$, $2$, and increasing $P=Q$. Each experiment is repeated $10$ times for each parameter setting. Let $C_q$ be the measured computational cost in terms of clock time measured in seconds of the algorithm in each case for given $q$. Then, Figure~\ref{fig1a} is the boxplot of each $C_q$ over the $10$ experiments; Figure~\ref{fig1b} shows the ratios $C_1/C_0$, $C_2/C_1$ and $C_2/C_1$; and  Figure~\ref{fig1c} shows the decreasing \texttt{RMSE} with increasing $q$.} \label{fig1}
\end{figure}

Figure~\ref{fig2} summarizes the influence of $r$, increasing from $1$ to $50$, on the computational cost of Algorithm~\ref{FBHMRSVDAlgorithm} with increasing $q$ for an example matrix  $\bfX$ of size of $141 \times 141$. The computational cost in terms of time measured in seconds increases approximately linearly with $r$ for each choice of $q$ and again for larger $r$ the cost may double going from $q=0$ to $q=1$, with a somewhat smaller increase from $q=1$ to $q=2$. Because matrix $\bfT$ is assumed to contain significant low-rank features of  the regional anomalies, a small value of $r$ is required.

\begin{figure}
\subfigure{\includegraphics[width=.65\textwidth]{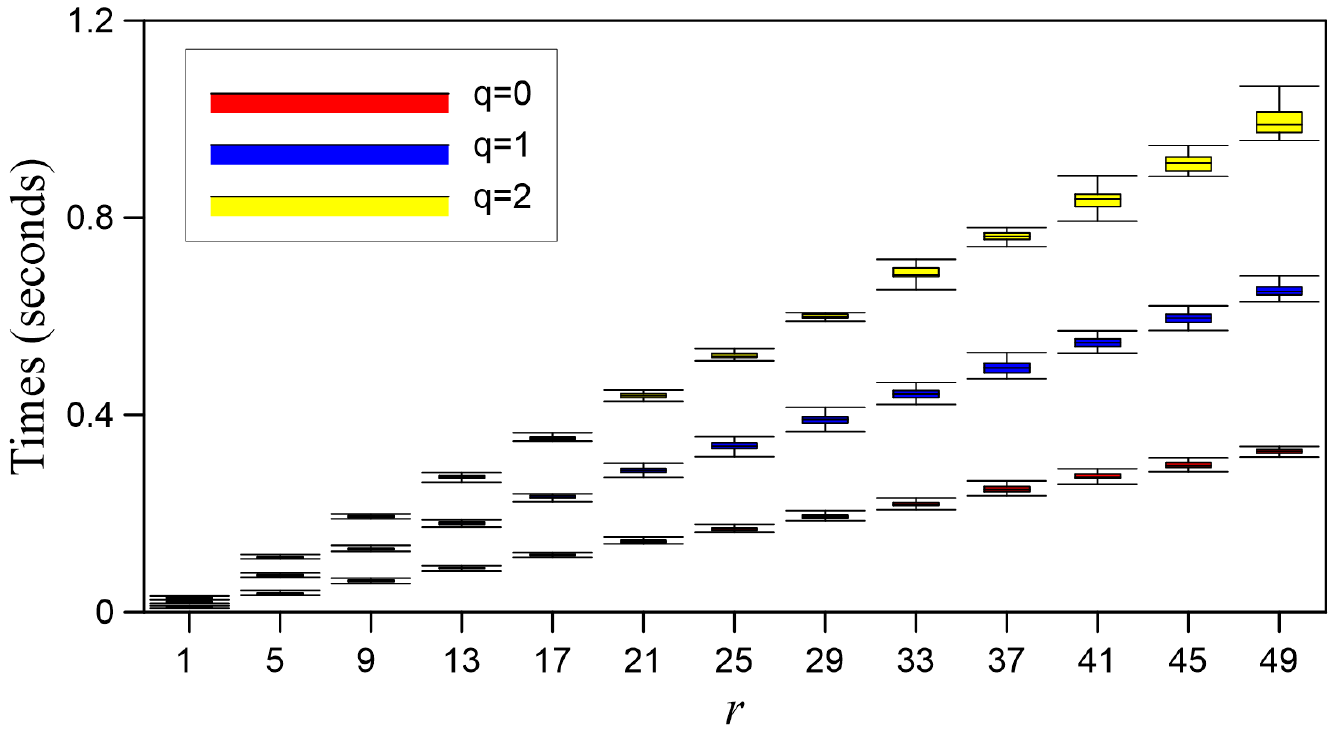}}
\caption{Experiments for Algorithm~\ref{FBHMRSVDAlgorithm}  with $q=0$, $1$, $2$ and increasing $r$, $r=1:50$. Each experiment is repeated $20$ times for each parameter setting. Computational times are reported in seconds.} \label{fig2}
\end{figure}

Table~\ref{tab2} and Figure~\ref{fig3} report on experiments that contrast the computational costs of Algorithm~\ref{FBHMRSVDAlgorithm}, both with and without use of fast matrix-matrix multiplication, \texttt{FBHMRSVD}, and \texttt{RSVD}, and for direct calculation using the partial \texttt{SVD}. Note that the \texttt{FBHMVM} can also be realized using the \texttt{1DFFT} \cite{lu2015fast}, and  thus, for comparison, the computational costs based on the \texttt{1DFFT} are also given in Table~\ref{tab2}. Here in Algorithm~\ref{FBHMRSVDAlgorithm} we use $r=10$ and $q=1$. Each experiment is performed $20$ times and  the median result  is reported in each case.  As reported in Table~\ref{tab2} it is immediate that  \texttt{FBHMRSVD} is most efficient for all sizes of $\bfX$. In addition, the computational costs using the  \texttt{2DFFT} are lower than those using the \texttt{1DFFT}. To examine the manner in which the cost improvement changes  with the size of $\bfX$, the ratios $T_{\mathtt{RSVD}}/T_{\mathtt{FBHMRSVD}}$ and $T_{\mathtt{SVD}}/T_{\mathtt{FBHMRSVD}}$ are shown in Figure~\ref{fig3}. Here $T_{\mathtt{RSVD}}$, $T_{\mathtt{FBHMRSVD}}$, and $T_{\mathtt{SVD}}$ denote the computational costs of \texttt{RSVD}, \texttt{FBHMRSVD}, and \texttt{SVD}, respectively. As  $\bfX$ increases in size, the computational cost of \texttt{FBHMRSVD} as compared to that of \texttt{RSVD} and \texttt{SVD} is relatively lower. Therefore, the reduction in computational cost is most significant when the size of $\bfX$ is large.

While these results suggest that the computational costs increase monotonically with increasing size of $\bfX$, we note that this may not always be observed. In particular, our code is implemented in MATLAB and uses the builtin MATLAB functions for the FFT and inverse FFT. But MATLAB has a mechanism to chose an optimal FFT algorithm dependent on the size of the transform that is required. Then a non-monotonic increase in computational cost can occur.  We demonstrate this feature of the MATLAB FFT implementation in Appendix~\ref{FFT cost of MATLAB}. 

To investigate the performance for a different release of MATLAB and compute environment, we also run a test with this configuration Intel(R) Core (TM) i7-6500U CPU @ 2.50GHz 2.60Hz; the memory is 8.00GB; the release of MATLAB is 2016a. The \texttt{SVD} and \texttt{RSVD} calculations are out of memory at $206\times 206$ and $241\times 241$. Thus, an advantage of our method is that it can be widely implemented on general computers.

\begin{table}
\caption{Comparisons of the computational times for Algorithm~\ref{FBHMRSVDAlgorithm}, both with and without use of fast matrix-matrix multiplication, \texttt{FBHMRSVD}, and \texttt{RSVD}, and direct calculation using the partial \texttt{SVD}. $/$ denotes that either the computational time is too high to perform the experiment, or an ``out of memory" error is reported.}\label{tab2}
\begin{tabular}{c  c | c c c c}
\hline
\multicolumn{2}{c}{Matrix sizes}&\multicolumn{4}{c}{Time (seconds)}\\ \hline
$\bfX$&$\bfT$&$\mathtt{FBHMRSVD (2DFFT)}$&$\mathtt{FBHMRSVD (1DFFT)}$&$\mathtt{RSVD}$&$\mathtt{SVD}$ $(r=10)$\\  \hline
$51 \times 51$& $676 \times 676$&$0.028$&$0.034$&$0.037$&$0.063$\\
$81 \times 81$& $1681 \times 1681$&$0.040$&$0.050$&$0.15$&$0.19$\\
$115 \times 115$& $3364 \times 3364$&$0.12$&$0.13$&$0.63$&$0.80$\\
$141 \times 141$& $5041 \times 5041$&$0.15$&$0.23$&$1.16$&$1.43$\\
$171 \times 171$& $7396 \times 7396$&$0.16$&$0.26$&$1.96$&$2.90$\\
$201 \times 201$& $10201 \times 10201$&$0.29$&$0.60$&$3.57$&$5.30$\\
$311 \times 311$& $22801 \times 22801$&$0.63$&$0.70$&$16.89$&$34.51$\\
$401 \times 401$& $40401 \times 40401$&$0.86$&$1.09$&$44.00$&$90.21$\\
$601 \times 601$& $90601 \times 90601$&$1.35$&$2.02$&$227.00$&$463.49$\\
$1001 \times 1001$& $251001 \times 251001$&$2.66$&$4.07$&$/$&$/$\\
$2001 \times 2001$& $1002001 \times 1002001$&$13.53$&$16.54$&$/$&$/$\\
\hline
\end{tabular}
\end{table}

\begin{figure}
\subfigure{\includegraphics[width=.65\textwidth]{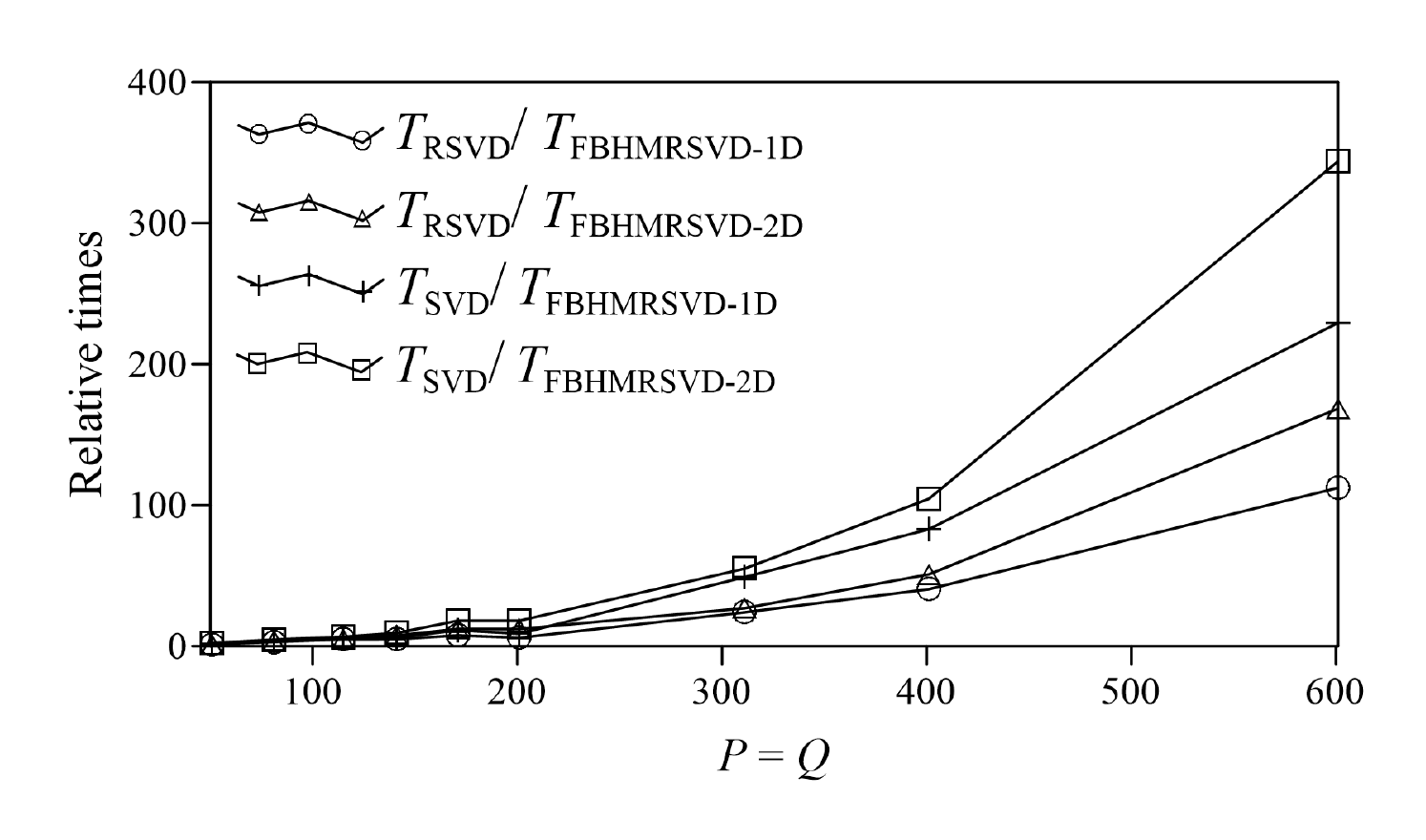}}\\
\caption{The comparative computational cost of obtaining a rank $r=10$ approximation for matrices of increasing size using Algorithm~\ref{FBHMRSVDAlgorithm} with $q=1$ both with and without use of fast matrix-matrix multiplication, \texttt{FBHMRSVD}, and \texttt{RSVD}, respectively, and compared with the direct use of the \texttt{SVD}. Each experiment is performed $20$ times and the ratios are reported for the median result in each case. } \label{fig3}
\end{figure}


\section{Fast non-convex low-rank matrix decomposition  potential field separation: \texttt{FNCLRMD\_PFS}}\label{FNCLRMDPFSMETHOD}
\subsection{Methodology}
We suppose that the total field data matrix of size $P \times Q$ is 
\bigskip
\begin{eqnarray*}
\bfX=\bfXD+\bfXS,
\end{eqnarray*}
\bigskip
where the gridded data matrices of the regional  and residual anomalies are denoted by $\bfXD$ and $\bfXS$,  respectively.  
Practically, $\bfXD$ and $\bfXS$ are unknown and  the objective of   potential field data separation is their estimation given $\bfX$. This means that the block Hankel matrix $\bfT=\mathcal{H}(\bfX)$ represents $\bfXD$ and $\bfXS$, separately,
\bigskip
\begin{eqnarray*}
\bfT=\bfTD+\bfTS = \mathcal{H}(\bfXD)+ \mathcal{H}(\bfXS).
\end{eqnarray*}
\bigskip
Because $\bfTD$ is assumed to have low rank and $\bfTS$ is assumed to be sparse, for which the derivations are given in   \cite{zhu2019low},  the separation can be achieved by solving the optimization problem
\bigskip
\begin{eqnarray}\label{PRCA optimization}
\min\:\{\rank(\bfTD),\|\bfTS\|_0\} \quad \subjectto\:\bfT=\bfTD+\bfTS.
\end{eqnarray}
\bigskip
The algorithm \texttt{LRMD\_PFS} introduced in \cite{zhu2019low} uses a convex method  to solve the optimization problem in  \eqref{PRCA optimization} by transforming to the optimization problem,
\bigskip
\begin{eqnarray}\label{NPRCA optimization}
\min\:\|(\bfTD)\|_*+\alpha\|\bfTS\|_1,\:\:\:\subjectto\:\bfT=\bfTD+\bfTS, 
\end{eqnarray}
\bigskip
where $\alpha>0$ denotes a weighting parameter. Because $\bfT$ is generally large, the solution of  \eqref{NPRCA optimization} is computationally demanding in terms of flops and memory. The \texttt{Altproj} Algorithm   \cite{netrapalli2014non} to solve \eqref{PRCA optimization} is, however, non-convex and proceeds by alternately updating $\bfTS$ by  projecting $\bfT-\bfTD$ onto the set of sparse matrices, and $\bfTD$  by projecting $\bfT-\bfTS$ onto the set of low-rank matrices. At each step the partial \texttt{SVD} of $\bfT$ is required. Thus, it is ideal to implement the \texttt{Altproj} Algorithm using the \texttt{FBHMRSVD} Algorithm~\ref{FBHMRSVDAlgorithm} for all estimates of the partial \texttt{SVD}. 
The solution of \eqref{PRCA optimization} with the application of the \texttt{Altproj} Algorithm combined with Algorithm~\ref{FBHMRSVDAlgorithm} is detailed in Algorithm~\ref{FNCLRMD_PFSAlgorithm}, and  is denoted by
\bigskip
\begin{eqnarray*}
[\bfXD^{*} ,\bfXS^{*}]=\mathtt{FNCLRMD\_PFS}(\bfX,K,\hat{K},r^{*},\beta,M, \epsilon).
\end{eqnarray*}
\bigskip
Here $r^{*}$, $\beta$, $M$, and $\epsilon$ are desired rank, thresholding parameter, an iteration parameter, and a convergence tolerance respectively.

\begin{algorithm}
\caption{Fast non-convex low-rank matrix decomposition  potential field separation: \\ $[\bfXD^{*} ,\bfXS^{*}]=\mathtt{FNCLRMD\_PFS}(\bfX,K,\hat{K},r^{*},\beta, M,\epsilon).$}\label{FNCLRMD_PFSAlgorithm}
\begin{algorithmic}[1]
\STATE \textbf{Input:} potential field data matrix $\bfX \in \RPQ$; parameter $K$ and $\hat{K}$; desired rank $r^{*}$; thresholding parameter $\beta$; iteration parameter $M$; convergence tolerance $\epsilon$.
\STATE \textbf{Definition:} $\sigma_{j}(\bfM)$ denotes the $j$th largest singular value of $\bfM$; $\mathtt{P}_{k}(\bfH)$ denotes the best rank $k$ approximation of $\bfH$; $\mathtt{HT}_{\zeta}(\bfH)$ denotes hard-thresholding applied to $\bfH$ such that entries with absolute values less than $\zeta$ are set to $0$; $\mathtt{IP}(\bfH)$ denotes inverse projection for recovering the matrix from its trajectory matrix by averaging the counter diagonal of each one of its blocks \cite{golyandina2007filtering}.
\STATE $[\bfU^{(0)},\bfSigma^{(0)},\bfV^{(0)}]=\mathtt{FBHMRSVD}(\bfX,K,\hat{K},1,1,1)$,
\STATE $\zeta_0=\beta \bfSigma^{(0)}(1,1)$.
\STATE $\bfXD^{(0)}=0$; $\bfXS^{(0)}=\mathtt{HT}_{\zeta_0}(\bfX-\bfXD)$.
\FOR {$k=1$ to $r^{*}$}
\FOR {$t=0$ to $M$}
\STATE $[\bfU^{(t+1)},\bfSigma^{(t+1)},\bfV^{(t+1)}]=\mathtt{FBHMRSVD}(\bfX-\bfXD^{(t)},K,\hat{K},k+1,k+1,1)$.
\STATE $\zeta=\beta(\bfSigma^{(t+1)}(k+1,k+1)+(\frac{1}{2})^{t} \bfSigma^{(t+1)}(k,k))$.
\STATE $\bfXD^{(t+1)}=\mathtt{IP}(\mathtt{P}_{k}(\bfX-\bfXD^{(t)}))$, where $\mathtt{P}_{k}(\bfX-\bfXD^{(t)})=\bfU_{k}^{(t+1)}\bfSigma_{k}^{(t+1)}(\bfV_{k}^{(t+1)})^T$.
\STATE $\bfXS^{(t+1)}=\mathtt{HT}_{\zeta}(\bfX-\bfXD^{(T+1)})$.
\IF {$\lVert\bfXD^{(t+1)}-\bfXD^{(t)}\rVert_{2}<\epsilon$}
\STATE \textbf{break} 
\ELSE
\STATE $\bfXS^{(0)}=\bfXS^{(t)}$.
\ENDIF
\ENDFOR
\ENDFOR
\STATE $\bfXD^*=\bfXD^{(t)}$, $\bfXS^*=\bfX-\bfXD^*$.
\STATE \textbf{Output:} $\bfXD^*$, $\bfXS^*$.
\end{algorithmic}
\end{algorithm}

\subsection{Parameter setting}
The quality of the solution of  \eqref{PRCA optimization} in terms of separating the regional and residual anomalies depends on the parameters $r^*$ and $\beta$. The default interval for the adjustment of $\beta$, $0< \beta <  1/\sqrt{\mathtt{max}(KL,\hat{K}\hat{L})}$ was recommended in \cite{zhu2019low}. Within this interval, experiments demonstrate that the results are consistent for a large subinterval. We will, see, however, that the quality of the separation is not very sensitive to the choice of $\beta$ and that the choice of $r^*$ is more significant. This is illustrated for synthetic geologic models for which the total field, regional anomaly, and residual anomaly are shown in Figures~\ref{fig4a}, \ref{fig4b}, \ref{fig5a}, and \ref{fig5b}, respectively. The parameters of these models, for which the matrices are of sizes $201 \times 201$, are detailed in Table~\ref{tab3}. 
Figure~\ref{fig6a} shows the \texttt{RMSE} when applying Algorithm~\ref{FNCLRMD_PFSAlgorithm} with $r^{*}=6$ and $10$ and different choices for $\beta$. While the \texttt{RMSE}s are relatively insensitive to  $\beta \in [0.0003,0.007]$, it is evident from  Figure~\ref{fig6b}, that the computational cost depends dramatically on the choice of $r^*$. This means the computational time is affected by $r^*$, but not $\beta$, and hence Algorithm~\ref{FNCLRMD_PFSAlgorithm} is relatively robust to the choice of $\beta$.

Now using $\beta \in [0.0003,0.007]$ as indicated from the previous experiment, the total field is separated for $r^*$ increasing from $1$ to $20$. 
 The \texttt{RMSE}s of the results are shown in Figure~\ref{fig6c}  and it is immediate  that the \texttt{RMSE} decreases rapidly for $r^*=1:4$,  but is relatively stable and independent of $r^*$ for $r^*>4$. On the other hand, it is clear from Figure~\ref{fig6d}, that the computational cost increases with increasing $r^*$. Thus there is a trade-off in terms of accuracy and computational cost in how $r^*$ is chosen.   Practically,  however, due to the low-rank features of the regional anomaly,  it is sufficient to take $r^*$ to be small, and generally not significantly larger than $10$.
 
Suitable separation of the anomalies is obtained using small values of the parameters $M$ and $\epsilon$. For all experiments  reported here we use  $M=10$ and $\epsilon=0$. Increasing $\epsilon$ to a small tolerance such as $10^{-3}$ or $10^{-7}$ will reduce the computational cost because the iteration will converge more quickly. Even with the chosen values, however,   \texttt{FNCLRMD\_PFS} is still much more efficient than \texttt{LRMD\_PFS}.
 
\begin{figure}
\subfigure[]{\label{fig4a}\includegraphics[width=.45\textwidth]{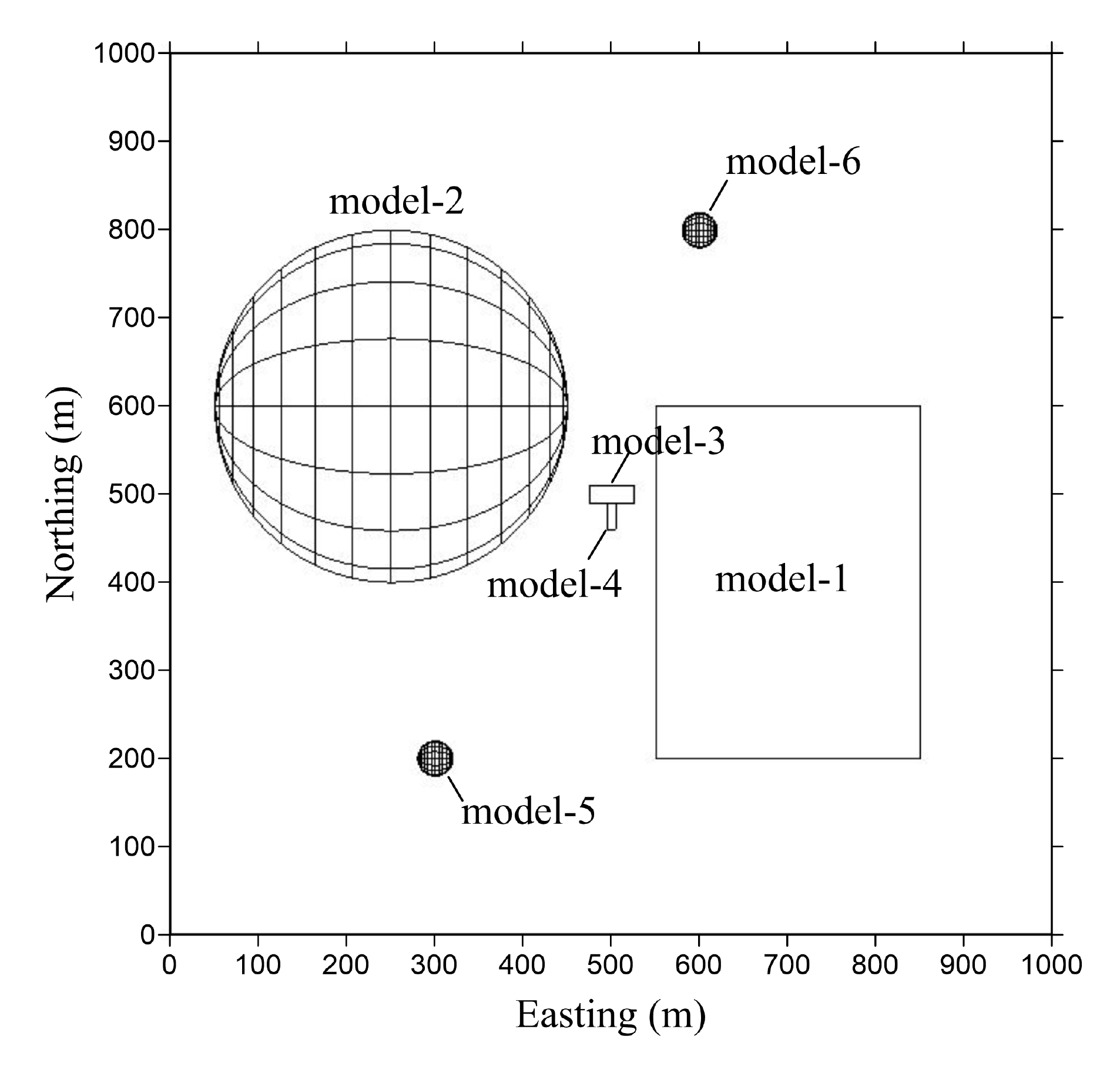}}
\subfigure[]{\label{fig4b}\includegraphics[width=.49\textwidth]{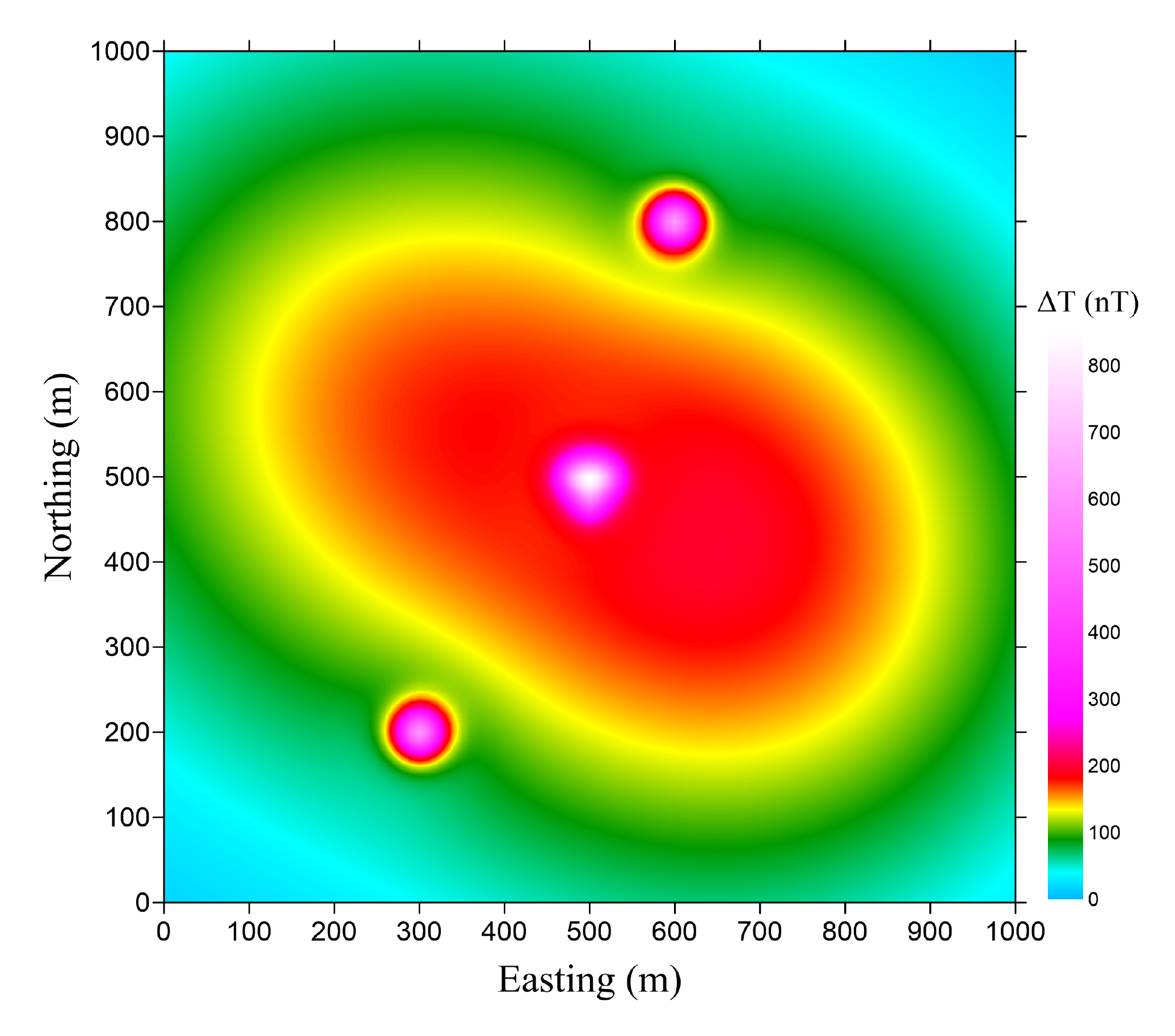}}
\caption{Figures~\ref{fig4a} and \ref{fig4b} are the geologic models and the forward magnetic field, respectively.} \label{fig4}
\end{figure}

\begin{table}
\caption{The parameters that define the geologic models in Figures~\ref{fig4a} and \ref{fig7a}.}\label{tab3}
\begin{tabular}{c c c c c c}
\hline
Geologic model&Shape&Central position & Model parameters & Density & Magnetization\\
& & &(length, width, depth extent)/radius&(g/cm$^3$)&(A/m)\\ \hline
model-$1$&Block&$(700,400,600)$&$(300,400,200)$&$0.5$&$8000$\\
model-$2$&sphere&$(250,600,700)$&$200$&$0.4$&$7000$\\
model-$3$&Block&$(500,500,40)$&$(50,20,40)$&$0.5$&$5000$\\
model-$4$&Block&$(500,475,40)$&$(10,30,40)$&$0.5$&$5000$\\
model-$5$&sphere&$(300,200,40)$&$20$& &$5000$\\
model-$6$&sphere&$(600,800,40)$&$20$& &$5000$\\
model-$7$&sphere&$(200,200,40)$&$20$&$0.7$&\\
model-$8$&Block&$(800,800,40)$&$(80,80,40)$&$0.5$&\\
\hline
\end{tabular}
\end{table}

\begin{figure}
\subfigure[]{\label{fig5a}\includegraphics[width=.49\textwidth]{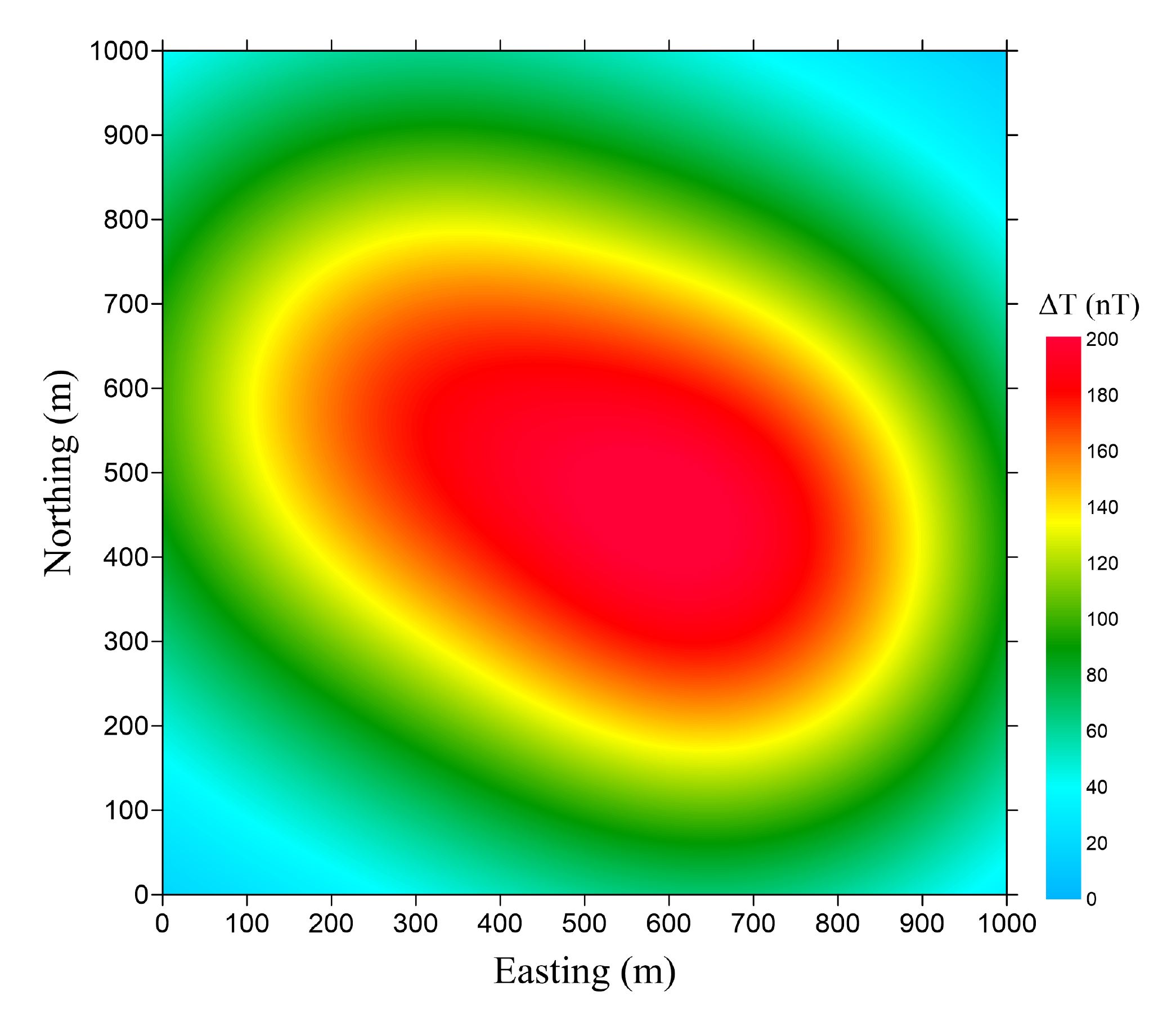}}
\subfigure[]{\label{fig5b}\includegraphics[width=.49\textwidth]{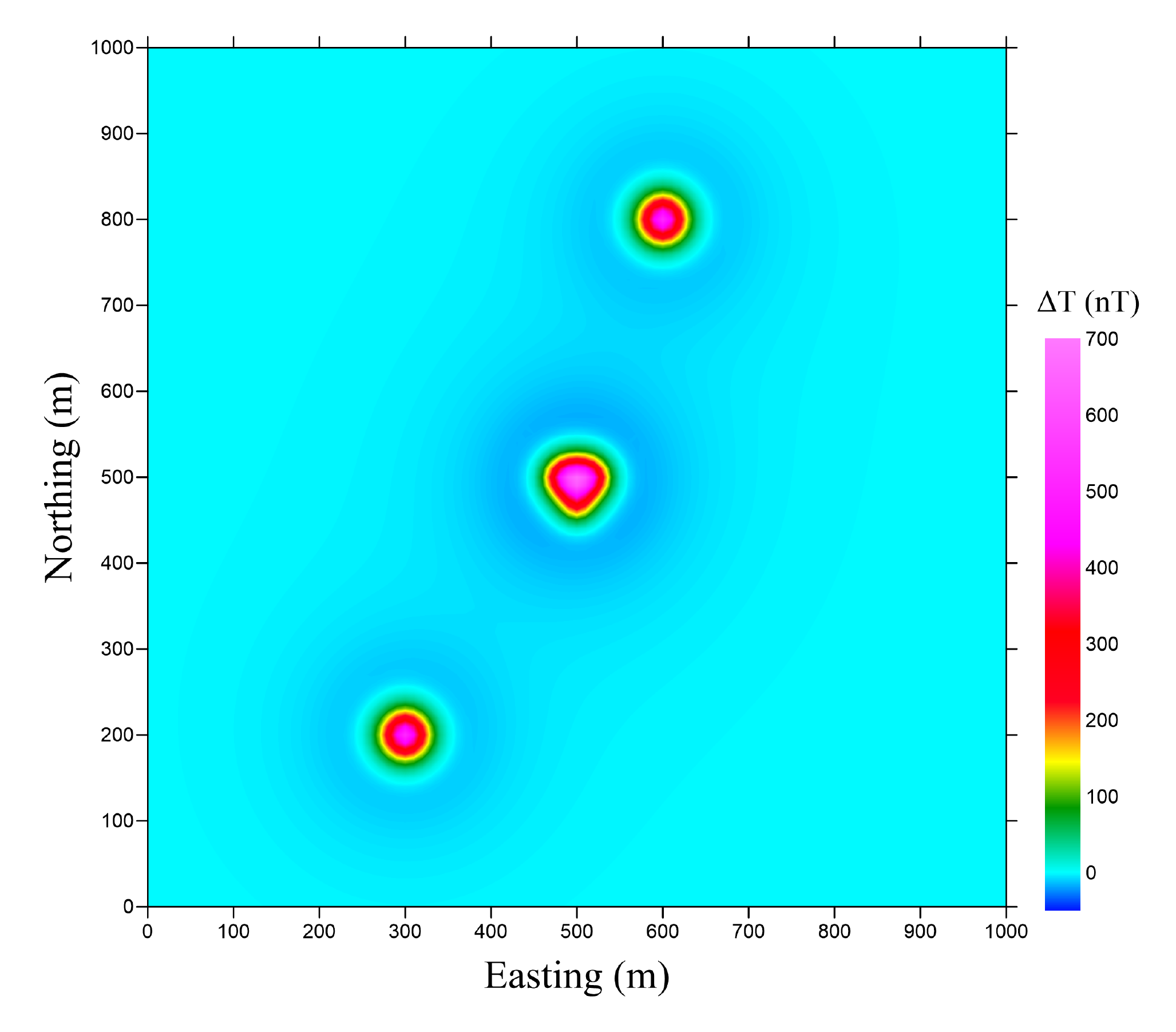}}\\
\subfigure[]{\label{fig5c}\includegraphics[width=.49\textwidth]{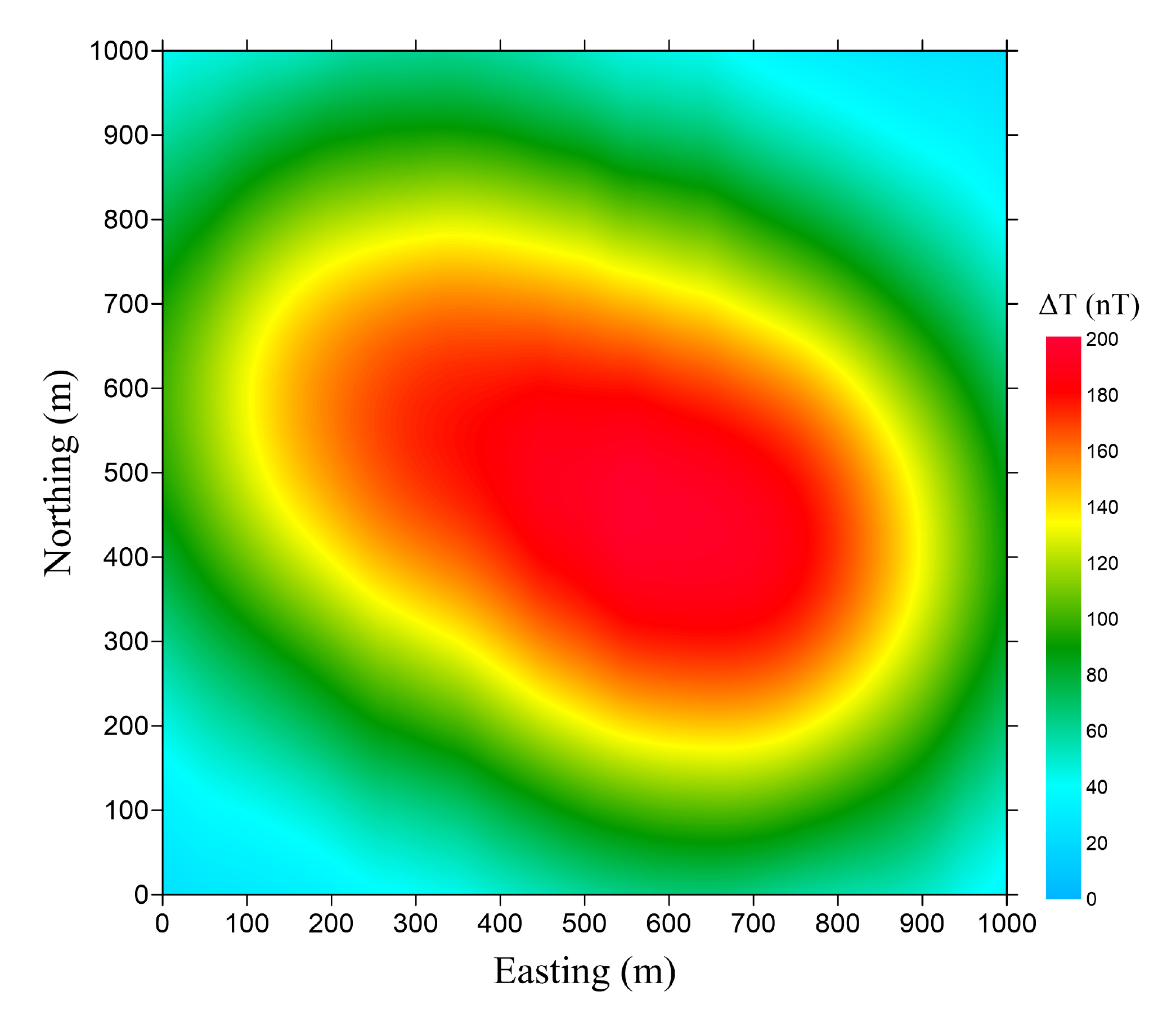}}
\subfigure[]{\label{fig5d}\includegraphics[width=.49\textwidth]{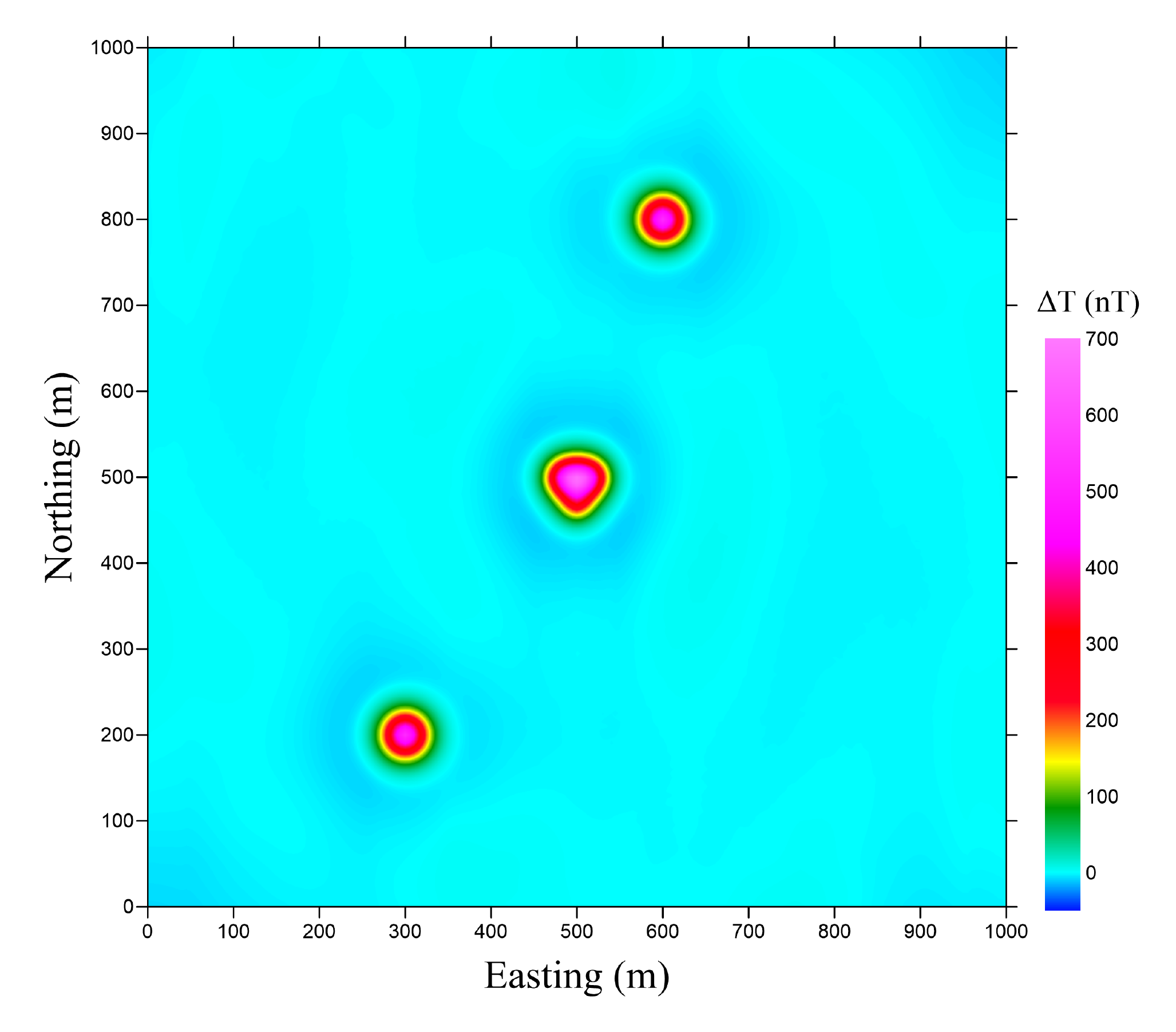}}\\
\subfigure[]{\label{fig5e}\includegraphics[width=.49\textwidth]{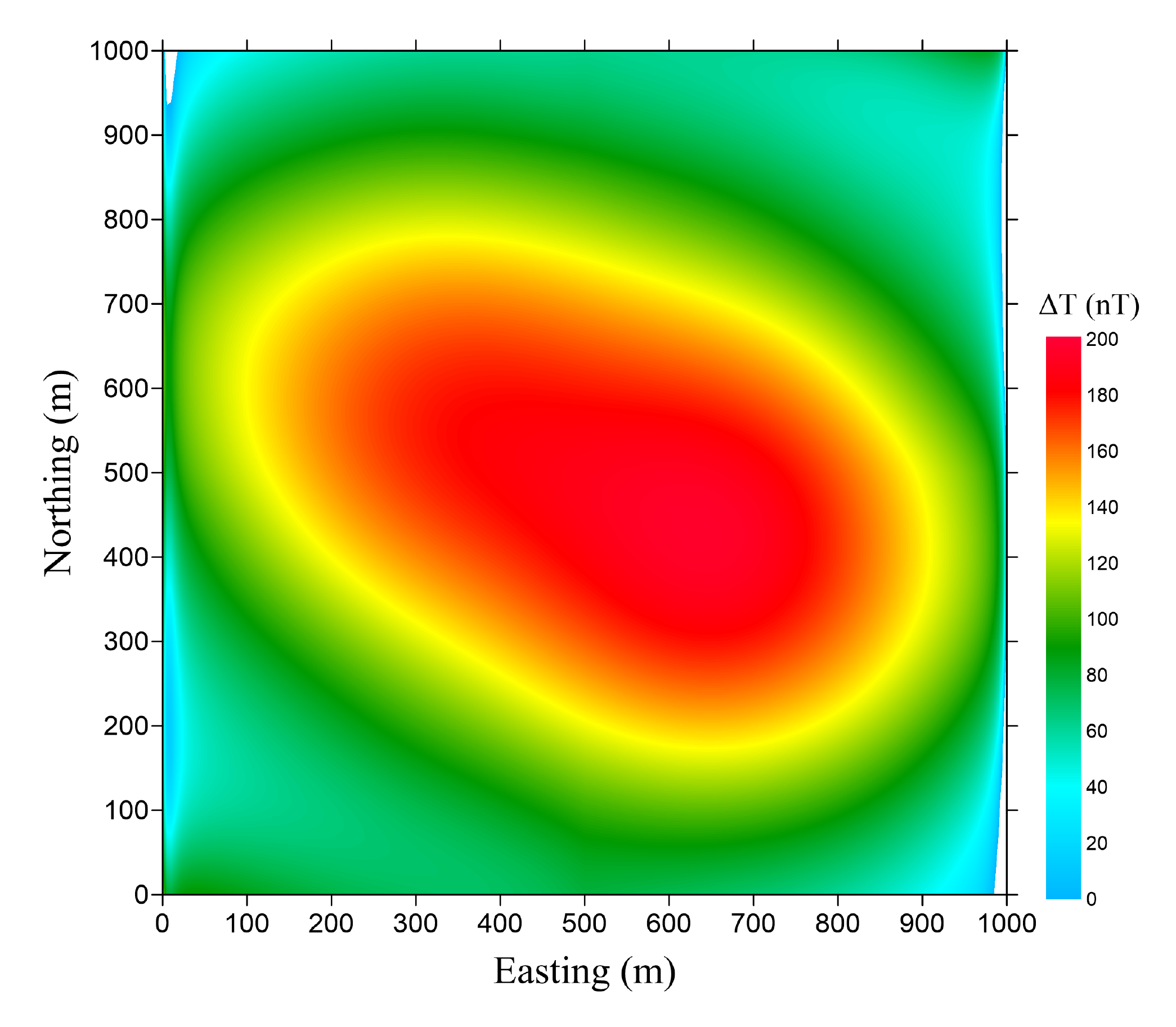}}
\subfigure[]{\label{fig5f}\includegraphics[width=.49\textwidth]{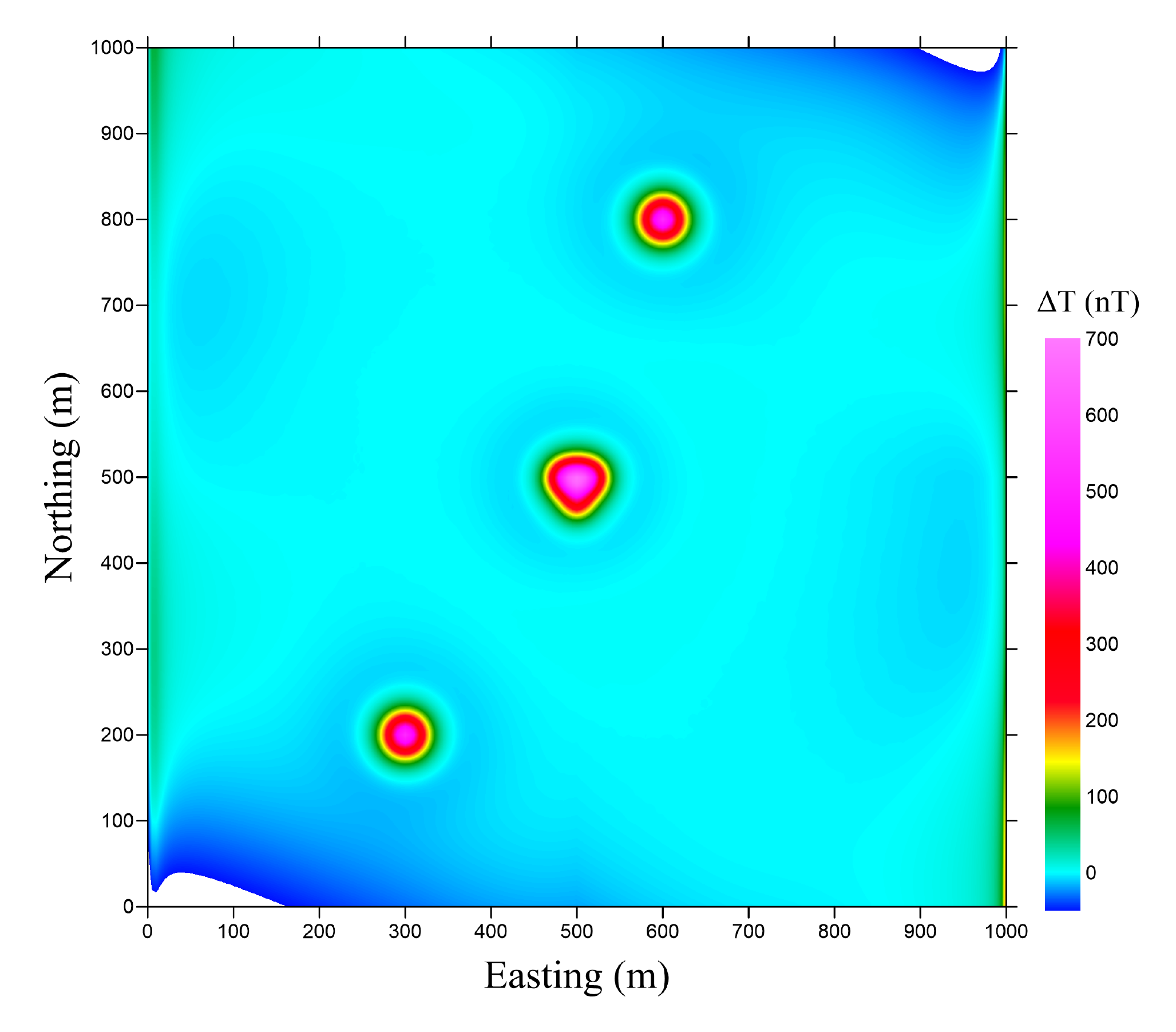}}
\caption{Figures~\ref{fig5a} and \ref{fig5b} are the synthetic regional and residual anomalies for the models in Figure~\ref{fig4}, respectively; Figures~\ref{fig5c} and \ref{fig5d}  are the separated regional and residual anomalies, respectively, for  data of size   $201 \times 201$ obtained using Algorithm~\ref{FNCLRMD_PFSAlgorithm} with $\beta = 0.0062$ and $r^* = 6$; Figures~\ref{fig5e} and \ref{fig5f} are the separated regional and residual anomalies,  respectively,  obtained using  \texttt{LRMD\_PFS} with $\alpha = 0.0005$.} \label{fig5}
\end{figure}

\begin{figure}
\subfigure[]{\label{fig6a}\includegraphics[width=.49\textwidth]{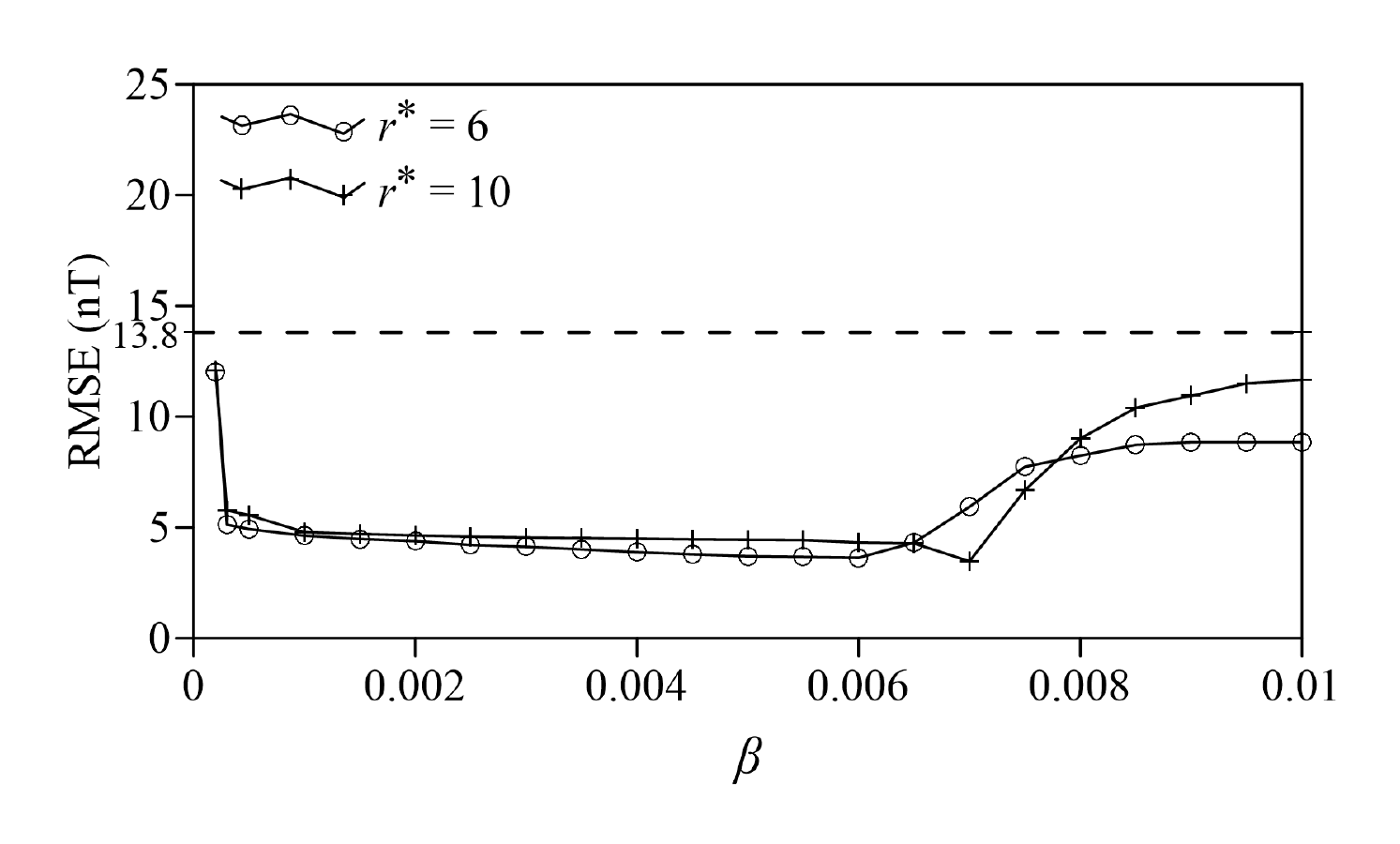}}
\subfigure[]{\label{fig6b}\includegraphics[width=.49\textwidth]{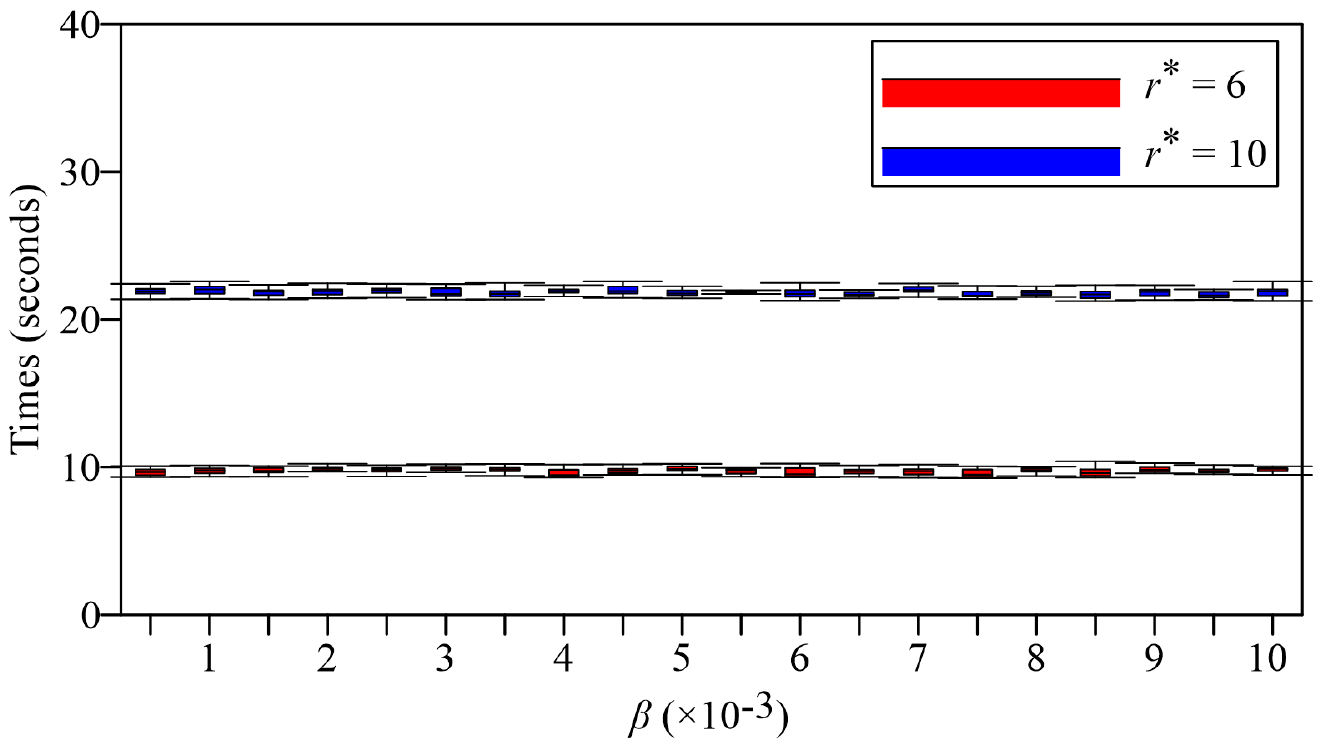}}\\
\subfigure[]{\label{fig6c}\includegraphics[width=.49\textwidth]{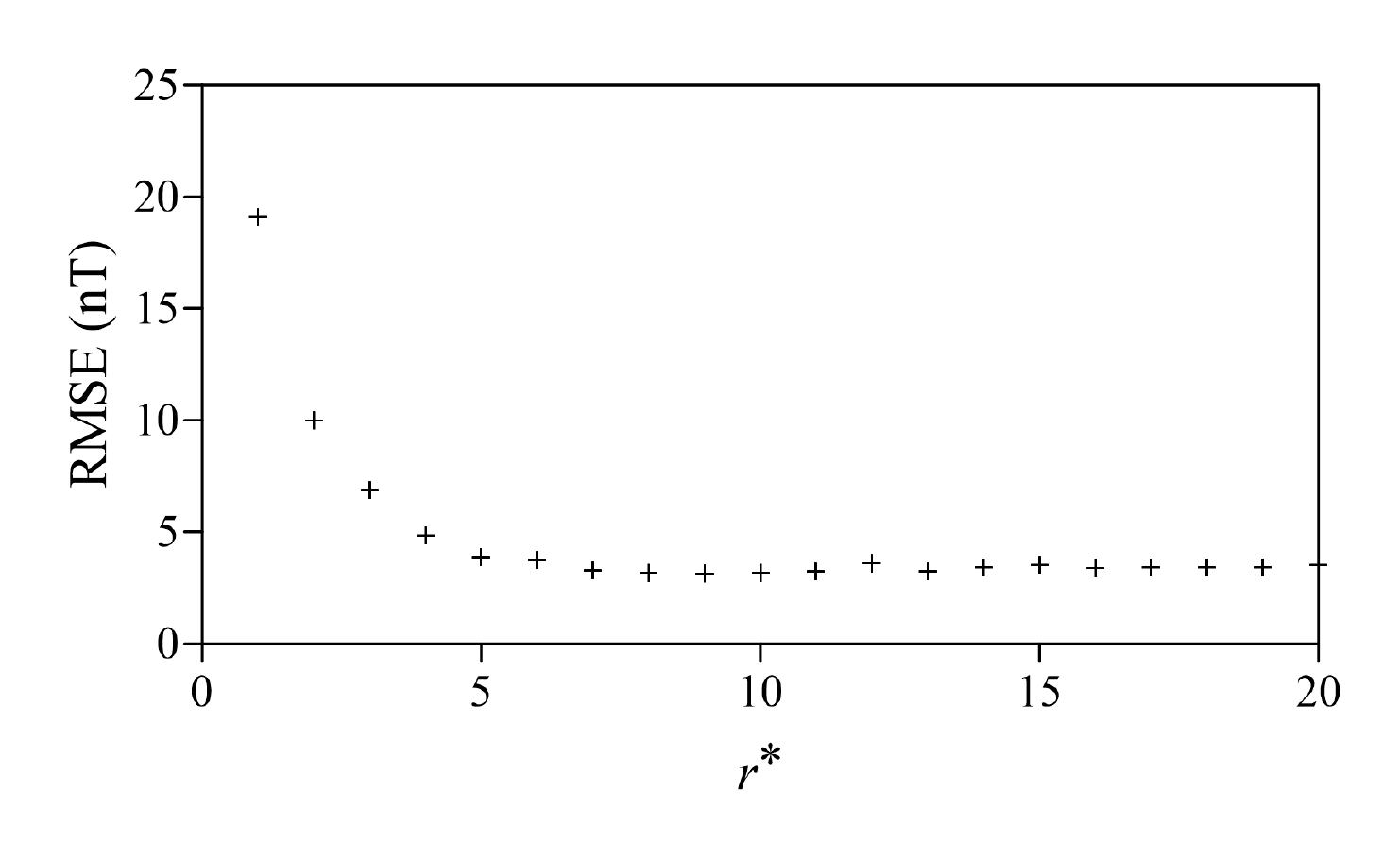}}
\subfigure[]{\label{fig6d}\includegraphics[width=.49\textwidth]{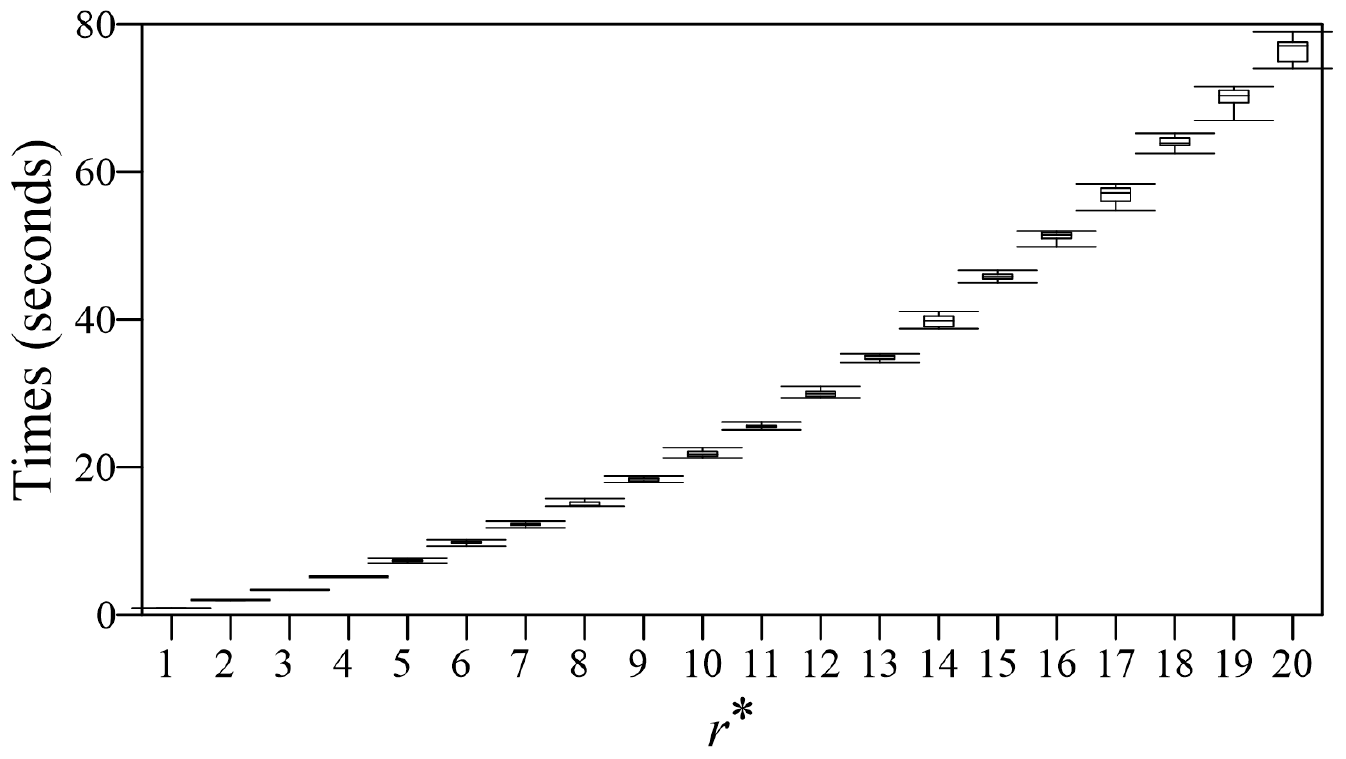}}\\
\caption{These results show tests of the parameters for Algorithm~\ref{FNCLRMD_PFSAlgorithm}. Figures~\ref{fig6a} and \ref{fig6b}  are the \texttt{RMSE} and the computational times for the separations of the data in Figure~\ref{fig4} with different $\beta$; Figures~\ref{fig6c} and \ref{fig6d} are the \texttt{RMSE} and the computational times of the separations of the data in Figure~\ref{fig4} with different $r^{*}$.} \label{fig6}
\end{figure}

\section{SYNTHETIC FIELD DATA EXPERIMENTS}\label{SyntheticResults}
\subsection{Experiment {1\label{exp:1}}: Magnetic Data}

The accuracy and the computational cost  of Algorithm~\ref{FNCLRMD_PFSAlgorithm}, dependent on $\beta$, for the gridded data matrices in Figure~\ref{fig4b}, for matrices of different sizes, is contrasted with  the results obtained using \texttt{LRMD\_PFS} dependent on $\alpha$.  The experiment is repeated for different choices of $\beta$ and $\alpha$ in the recommended intervals, and the result with the smallest \texttt{RMSE} is selected as the final result and reported in  Table~\ref{tab4}, with illustration in Figures~\ref{fig5}-\ref{fig6}. The \texttt{RMSE}s obtained using Algorithm~\ref{FNCLRMD_PFSAlgorithm} are   between $3.53$ and $3.78$ nT, while the \texttt{RMSE}s for the same experiments  using \texttt{LRMD\_PFS} are between $13.80$ and $16.39$ nT, hence demonstrating the higher accuracy of the new algorithm. It is more significant, however, that Algorithm~\ref{FNCLRMD_PFSAlgorithm} performs better than \texttt{LRMD\_PFS} with respect to computational cost in terms of computational time and memory demand. Moreover, in the given computational environment, it is not possible to obtain the data matrices of sizes much greater than $205 \times 205$ using \texttt{LRMD\_PFS}. In contrast,  it is possible to solve the problem for matrices of sizes $2001 \times 2001$ using Algorithm~\ref{FNCLRMD_PFSAlgorithm}.  For the smaller problem of size $201 \times 201$, Figures~\ref{fig5c} and \ref{fig5d} show the separated regional and residual anomalies using Algorithm~\ref{FNCLRMD_PFSAlgorithm}, while  Figures~\ref{fig5e} and \ref{fig5f} show the separated anomalies obtained using \texttt{LRMD\_PFS}. It can be seen from  Figures~\ref{fig5c} to \ref{fig5f}, that Algorithm~\ref{FNCLRMD_PFSAlgorithm}  performs well around the boundaries, but that the two methods are comparable in the central areas.

\begin{table}
\caption{Comparisons of the computational times of the \texttt{FBHMRSVD}, \texttt{RSVD}, and \texttt{SVD}. $/$ denotes that either the computational time is too high to perform the experiment, or an ``out of memory" error is reported.}\label{tab4}
\begin{tabular}{c  c | c c c c | c c c}
\hline
\multicolumn{2}{c}{Matrix sizes}&\multicolumn{4}{c}{\texttt{FNCLRMD\_PFS}}&\multicolumn{3}{c}{\texttt{LRMD\_PFS}}\\ \hline
$\bfX$&$\bfT$&$r^*$&$\beta$&\texttt{RMSE} (nT)&Times (s)&$\alpha$&\texttt{RMSE} (nT)&Times (s)\\  \hline
$141 \times 141$& $5041 \times 5041$&$6$&$0.013$&$3.53$&$4.93$&$0.001$&$15.10$&$128.28$\\
$171 \times 171$& $7396 \times 7396$&$6$&$0.0088$&$3.64$&$5.51$&$0.0008$&$14.46$&$446.20$\\
$201 \times 201$& $10201 \times 10201$&$6$&$0.0062$&$3.60$&$9.77$&$0.0005$&$13.80$&$1043.37$\\
$311 \times 311$& $22801 \times 22801$&$6$&$0.0026$&$3.61$&$23.94$&$/$&$/$&$/$\\
$401 \times 401$& $40401 \times 40401$&$6$&$0.0014$&$3.65$&$40.42$&$/$&$/$&$/$\\
$601 \times 601$& $90601 \times 90601$&$6$&$0.0007$&$3.65$&$87.21$&$/$&$/$&$/$\\
$1001 \times 1001$& $251001 \times 251001$&$6$&$0.0002$&$3.75$&$218.54$&$/$&$/$&$/$\\
$2001 \times 2001$& $1002001 \times 1002001$&$6$&$0.00002$&$4.22$&$1062.29$&$/$&$/$&$/$\\
\hline
\end{tabular}
\end{table}

\subsection{Experiment 2: Gravity Data\label{exp:2}}
For this experiment, the synthetic geologic models, the total field, the regional anomaly and the residual gravity anomaly are shown in Figures~\ref{fig7a}, \ref{fig7b}, \ref{fig8a}, and \ref{fig8b}, respectively. The parameters that define the models, all for data matrices of size $201 \times 201$,  are detailed in Table~\ref{tab3}, and the results are illustrated in Figure~\ref{fig8}. In contrast to Experiment \textbf{1} (in \ref{exp:1}),  the residual anomaly is generated for geologic models with different scales. 
In the application of Algorithm~\ref{FNCLRMD_PFSAlgorithm} for the separation of the data we set $r^*=6$ and $\beta=0.0005$. This yields a  \texttt{RMSE} of $0.0028$ mGal. In contrast the smallest \texttt{RMSE} using \texttt{LRMD\_PFS} is $0.017$ mGal and is obtained with $\alpha=0.0007$. Thus, Algorithm~\ref{FNCLRMD_PFSAlgorithm} yields a higher accuracy result. Moreover, the computational clock times are $46.47$ and $2249.69$ s, respectively. 
Hence, Algorithm~\ref{FNCLRMD_PFSAlgorithm} is much more efficient. The results of the separation by the two methods are shown in Figures~\ref{fig8c} to \ref{fig8f}. It can be seen from  Figures~\ref{fig8c} to \ref{fig8f}, that the obtained gravity values of the separated regional anomaly in the north-east  region using \texttt{LRMD\_PFS} are higher than the synthetic regional anomaly. Thus, not only is Algorithm~\ref{FNCLRMD_PFSAlgorithm} more efficient, the results are qualitatively better.

In summary, our experiments demonstrate that Algorithm~\ref{FNCLRMD_PFSAlgorithm} has higher accuracy and lower computational cost than the \texttt{LRMD\_PFS} for the separation of both magnetic and gravity data.

\begin{figure}
\subfigure[]{\label{fig7a}\includegraphics[width=.44\textwidth]{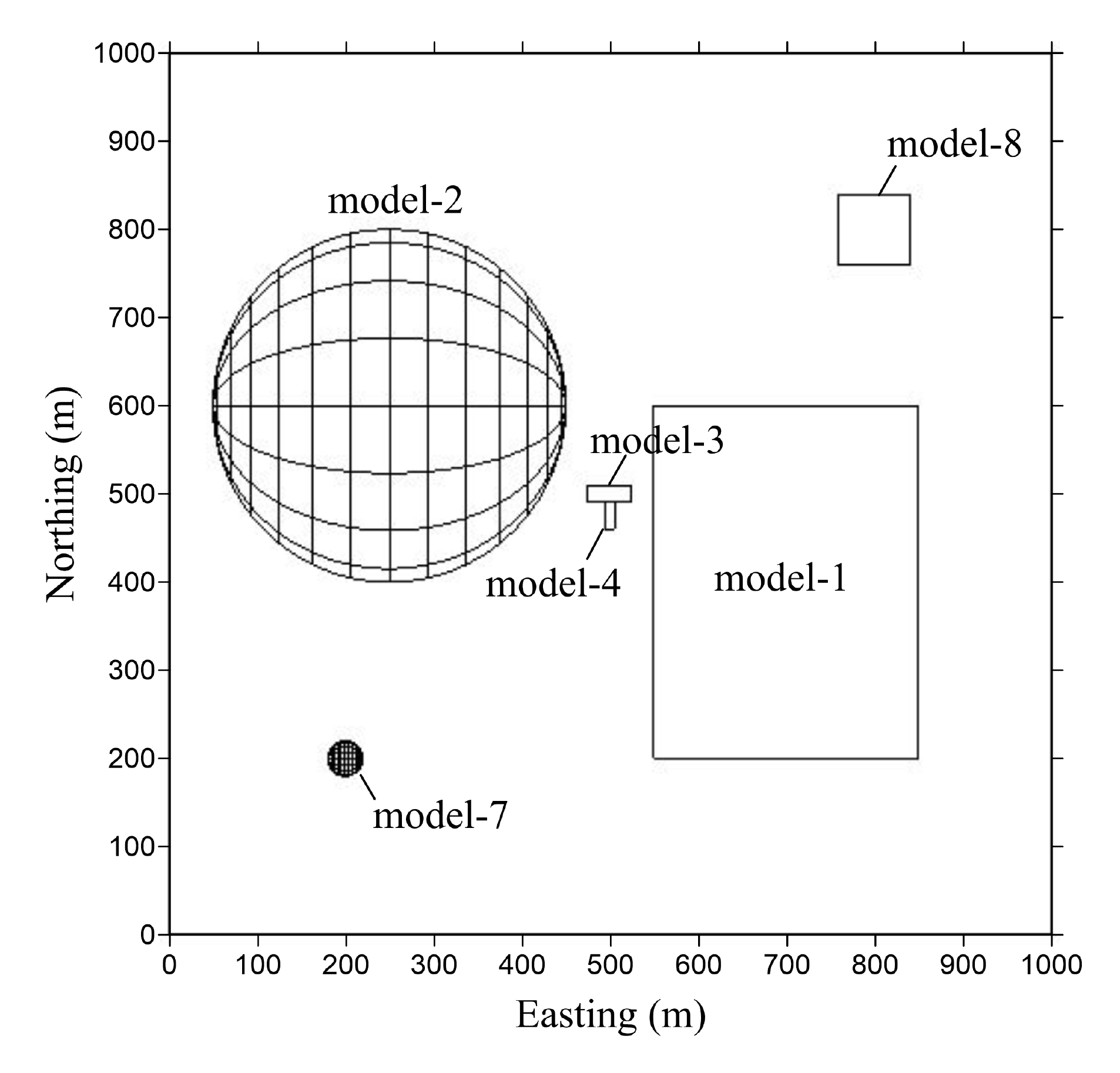}}
\subfigure[]{\label{fig7b}\includegraphics[width=.49\textwidth]{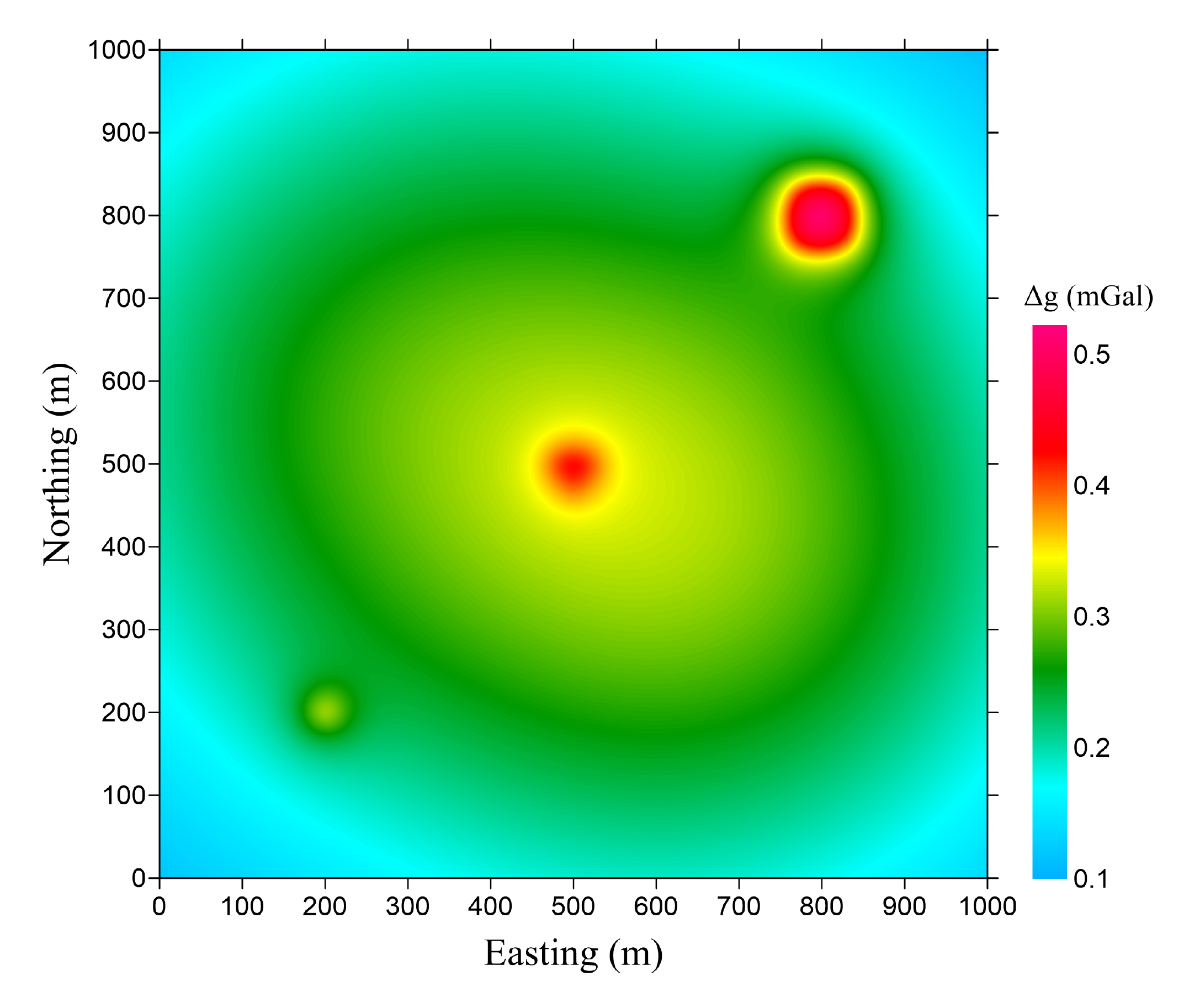}}\\
\caption{Figures~\ref{fig7a} and \ref{fig7b} are the geologic models and the forward gravity field, respectively.} \label{fig7}
\end{figure}

\begin{figure}
\subfigure[]{\label{fig8a}\includegraphics[width=.49\textwidth]{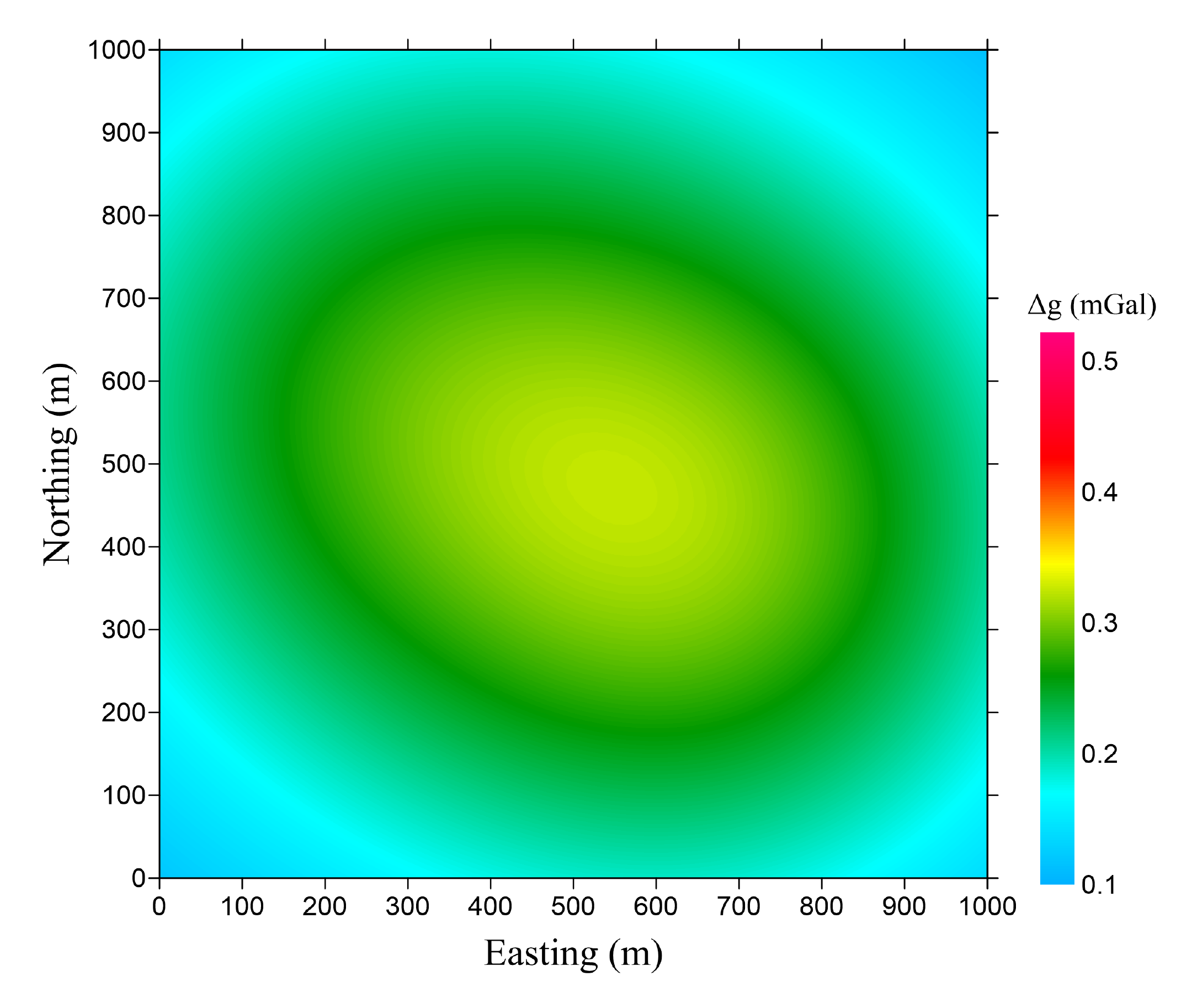}}
\subfigure[]{\label{fig8b}\includegraphics[width=.49\textwidth]{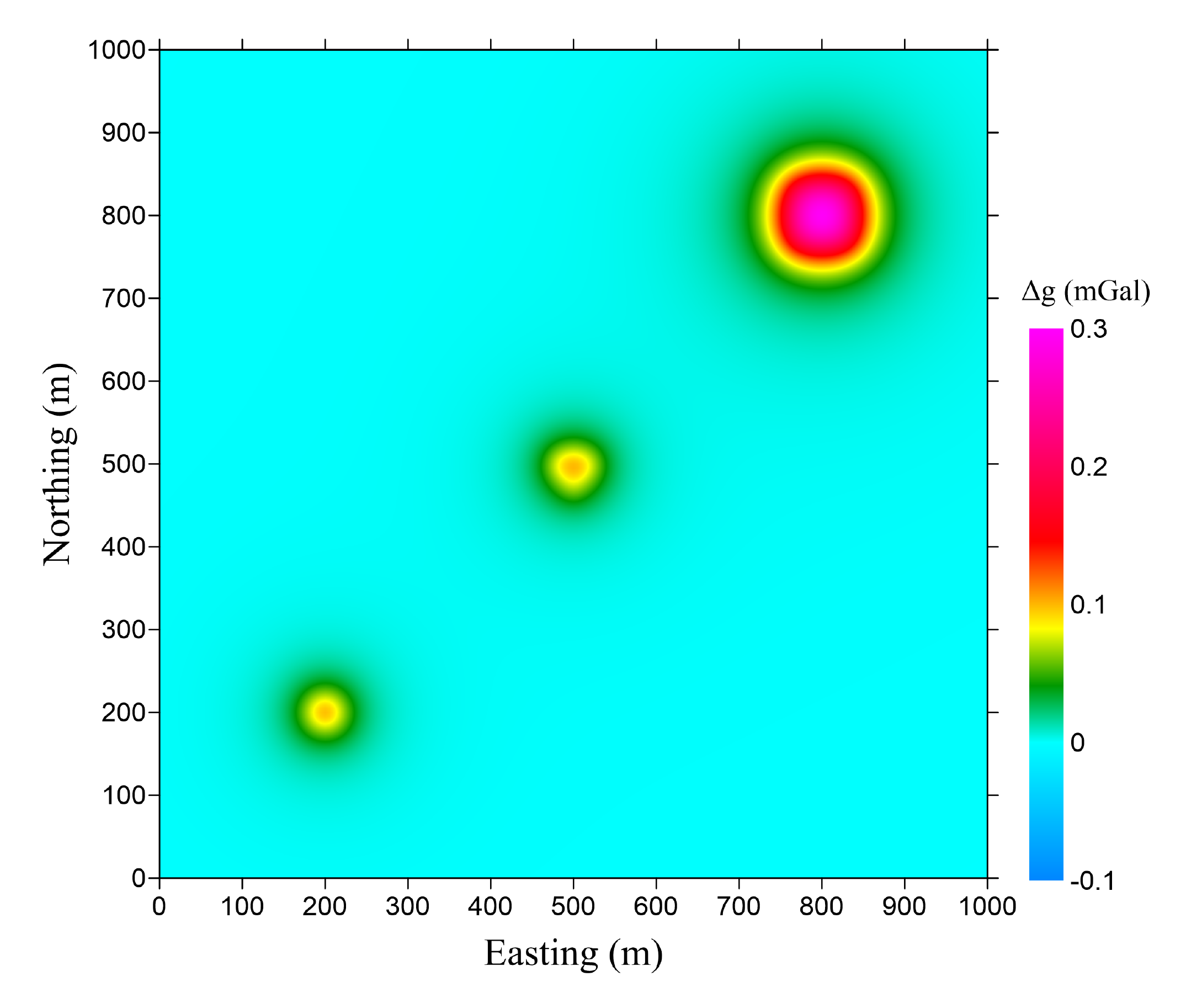}}\\
\subfigure[]{\label{fig8c}\includegraphics[width=.49\textwidth]{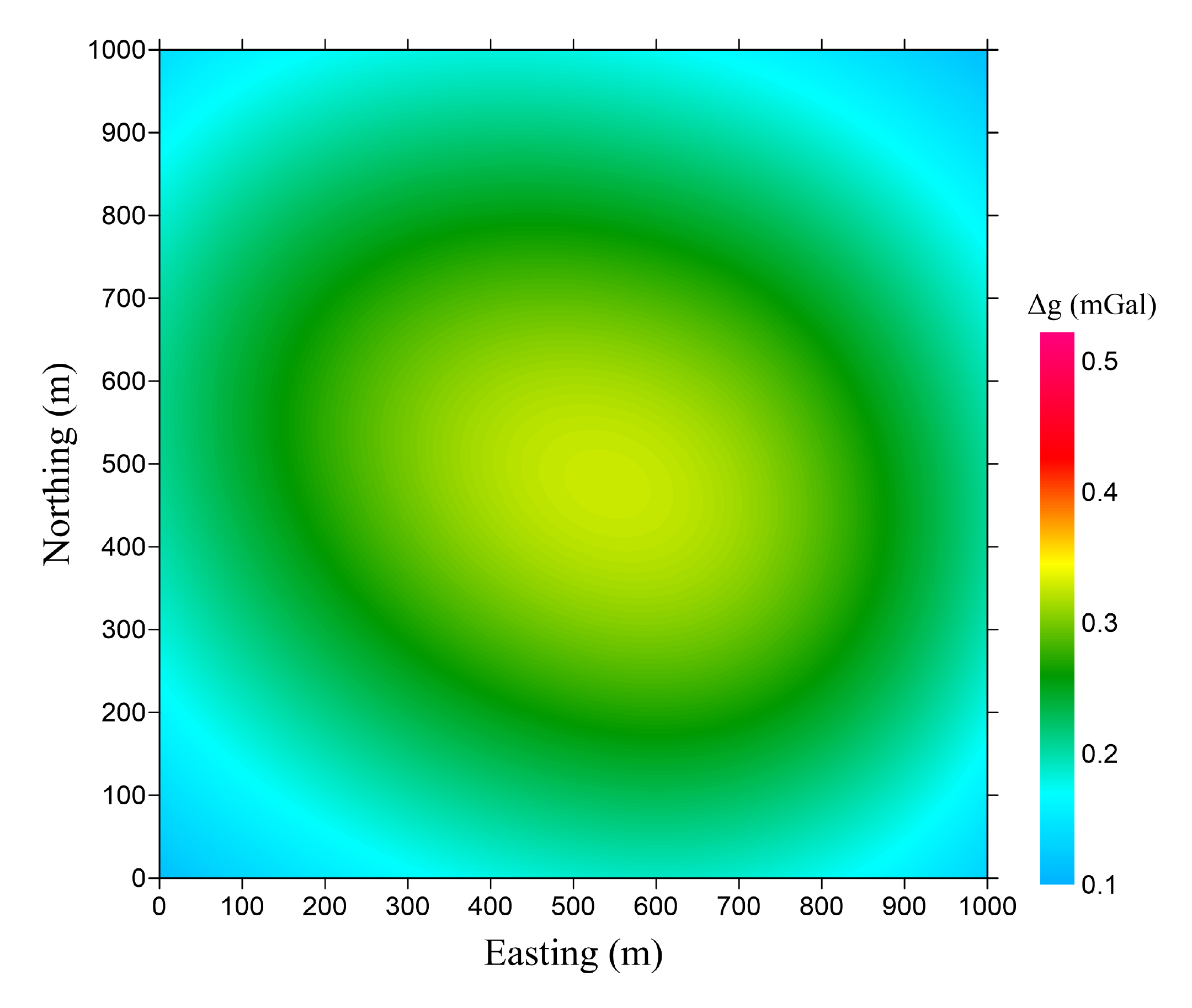}}
\subfigure[]{\label{fig8d}\includegraphics[width=.49\textwidth]{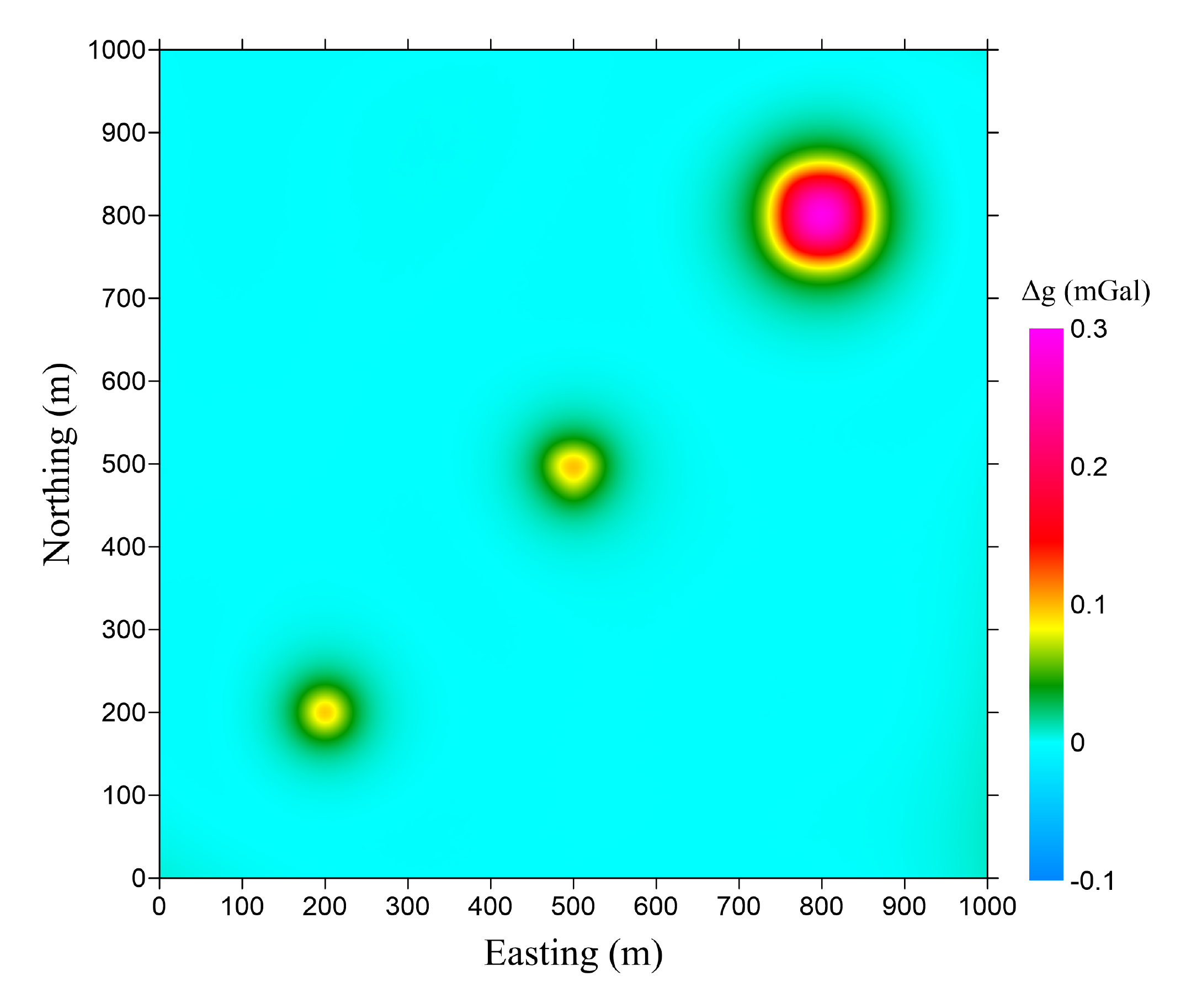}}\\
\subfigure[]{\label{fig8e}\includegraphics[width=.49\textwidth]{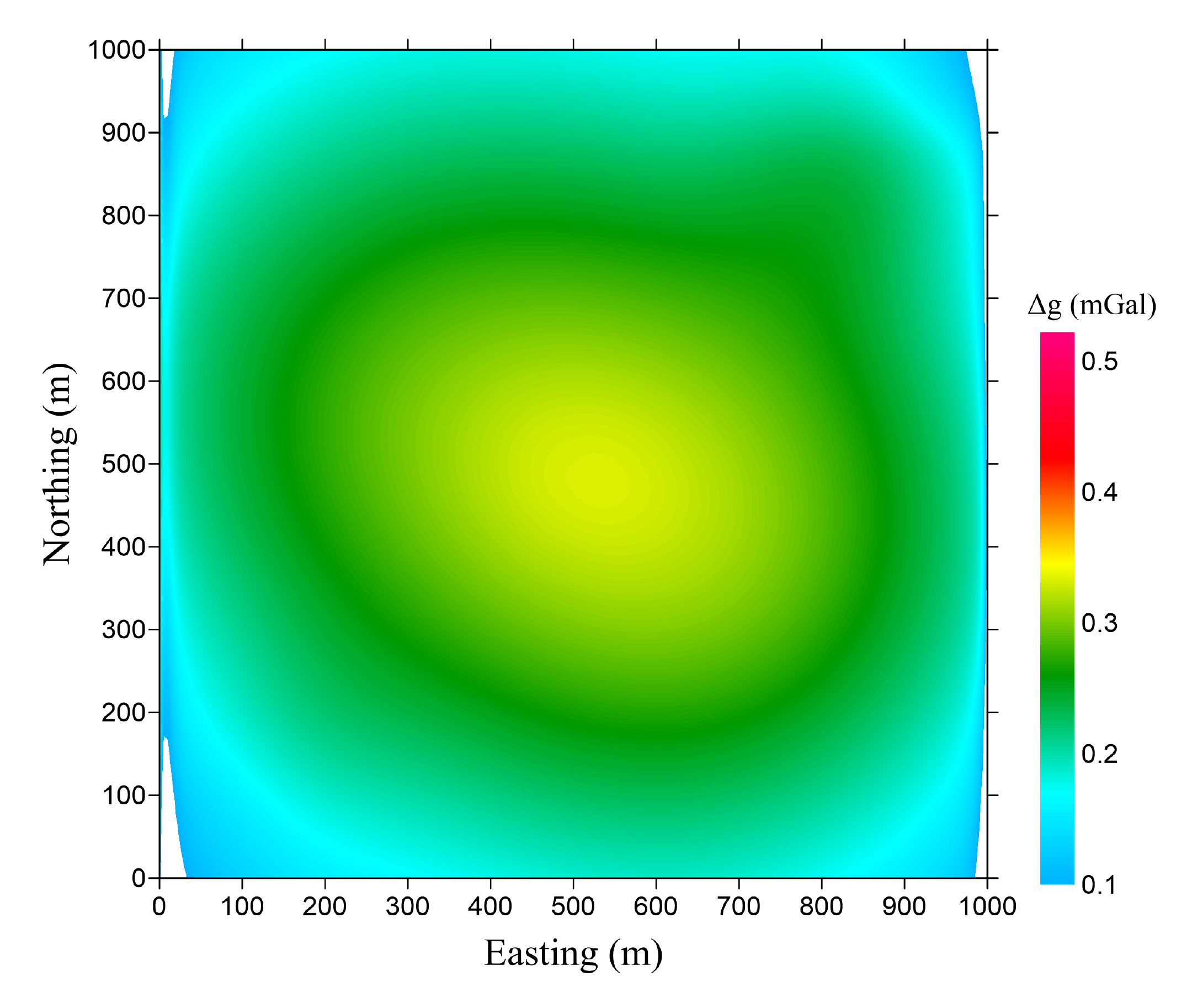}}
\subfigure[]{\label{fig8f}\includegraphics[width=.49\textwidth]{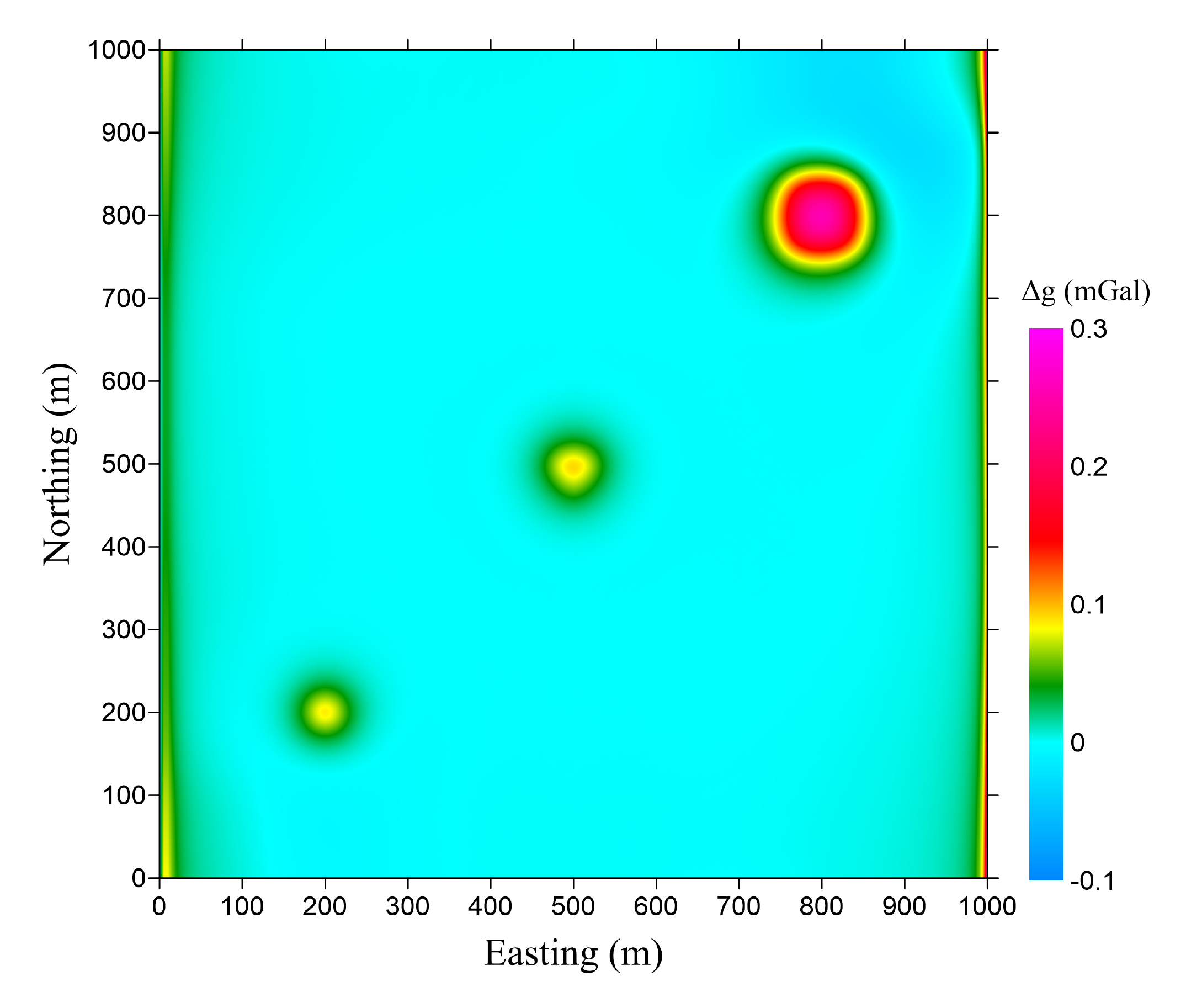}}
\caption{
Figures~\ref{fig8a} and \ref{fig8b} are the synthetic regional and residual anomalies for the models in Figure~\ref{fig7}, respectively; Figures~\ref{fig8c} and \ref{fig8d}  are the separated regional and residual anomalies, respectively, for  data of size   $201 \times 201$ obtained using Algorithm~\ref{FNCLRMD_PFSAlgorithm} with $\beta = 0.0005$ and $r^* = 6$; Figures~\ref{fig8e} and \ref{fig8f} are the separated regional and residual anomalies, respectively, obtained using  \texttt{LRMD\_PFS} with $\alpha = 0.0007$.} \label{fig8}
\end{figure}


\section{An Investigation of \texttt{FNCLRMD\_PFS} for a practical data set}\label{PracticalData}
The Tongling region is a good example of skarn deposits  lying in Anhui province of China. The Fenghuangshan copper deposit, which is a famous area in the Tongling region, is situated in the east central of the Middle-lower Yangtze metallogenic belt. The mineral deposits are generally of  hydrothermal metasomatic type.  Thus, the ore bodies occur in the contact zones between  igneous rocks and  sedimentary rocks. Therefore, in order to predict the location of concealed ore bodies, the separation of anomalies produced by igneous rocks is required.

The study area has three types of rocks. These include sedimentary rocks, igneous rocks, and skarn (or ore body). The physical properties of the sedimentary rocks are medium densities and non-magnetizations, while   the igneous rocks are low density (with residual density $-0.1$ g/cm$^{3}$) and medium magnetization (with magnetic susceptibility  $0 \thicksim 3400 \times 10^{-6} \times 4 \pi$ SI). In contrast, the  skarn and ore bodies are  of high density (with residual density  $0.7$ g/cm$^{3}$) and strong magnetization (with magnetic susceptibility  larger than $10000 \times 10^{-6} \times 4 \pi$ SI). The difference in the density and magnetic properties of these different rocks makes it effective to study the igneous rocks and ore bodies through gravity and magnetic exploration. Our objective is to separate the combination of regional anomalies of low-gravity and high-magnetism that are produced by igneous rocks,  and the combination of local anomalies of high-gravity and high-magnetism produced by skarn and ore bodies. Thus providing a basis for inversion and interpretation.

The algorithm is applied for the separation of the anomalies in Figure~\ref{fig9}. Figure~\ref{fig9a} is a Bouguer gravity anomaly map. The Bouguer gravity anomalies in the study area are high in the north (about $12$mGal) and low in the south (about $3$mGal). The data matrix has size  $247 \times 257$. In separating the gravity field we use $r^*=10$ and $\beta=0.01$, yielding the separated regional and residual gravity anomalies  shown in Figures~\ref{fig10a} and \ref{fig10b}, respectively.

The reduce to the pole (\texttt{RTP}) magnetic anomaly is shown in Figure~\ref{fig9b}. There is a local high magnetic anomaly centered around Xinwuling. The size of the magnetic data matrix is $197 \times 199$.  In separating the \texttt{RTP} magnetic  field we use $r^*=10$ and $\beta=0.005$, yielding the separated regional and residual magnetic anomalies  shown in Figures~\ref{fig10c} and \ref{fig10d}, respectively.

\begin{figure}
\subfigure[]{\label{fig9a}\includegraphics[width=.49\textwidth]{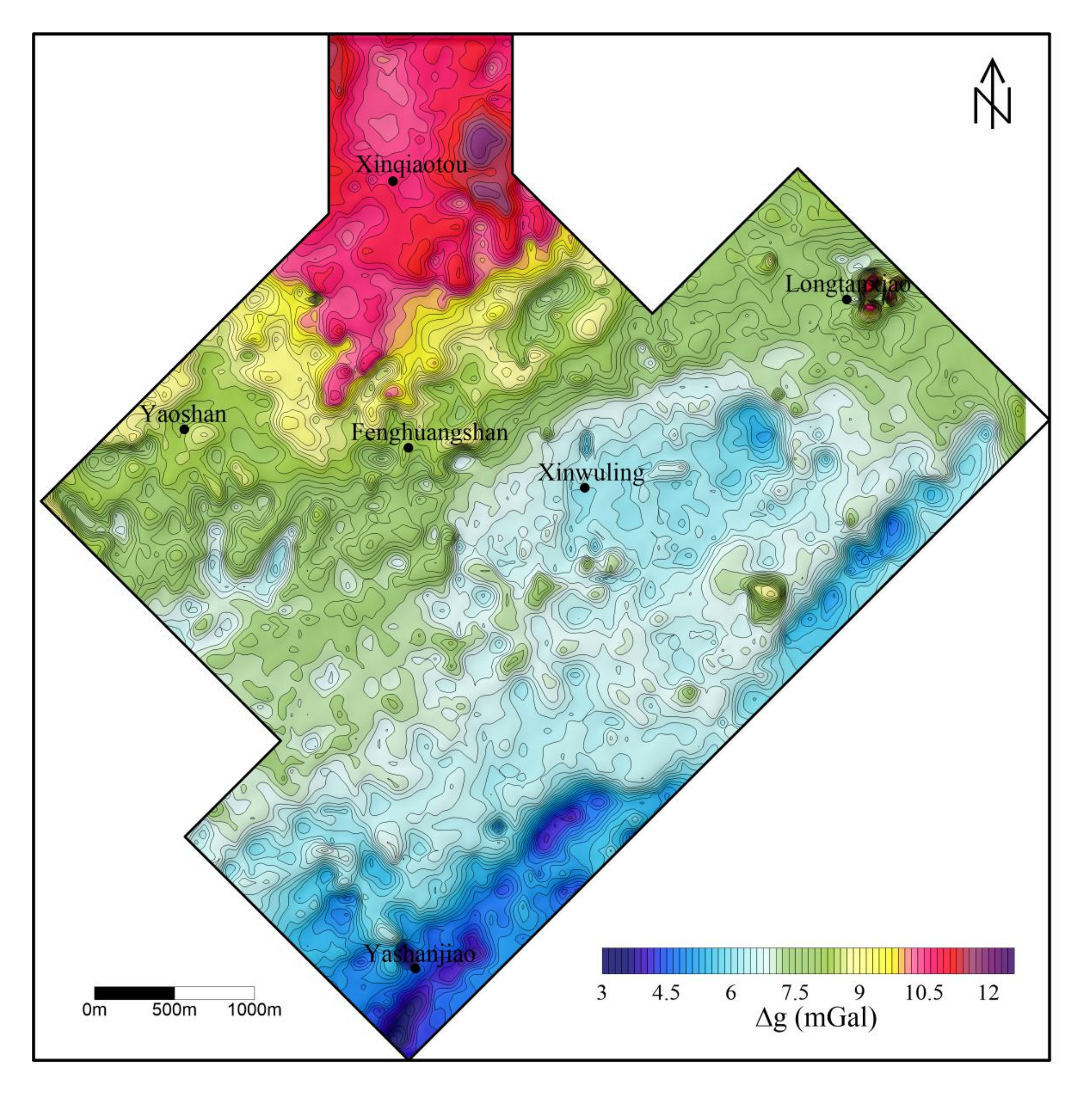}}
\subfigure[]{\label{fig9b}\includegraphics[width=.49\textwidth]{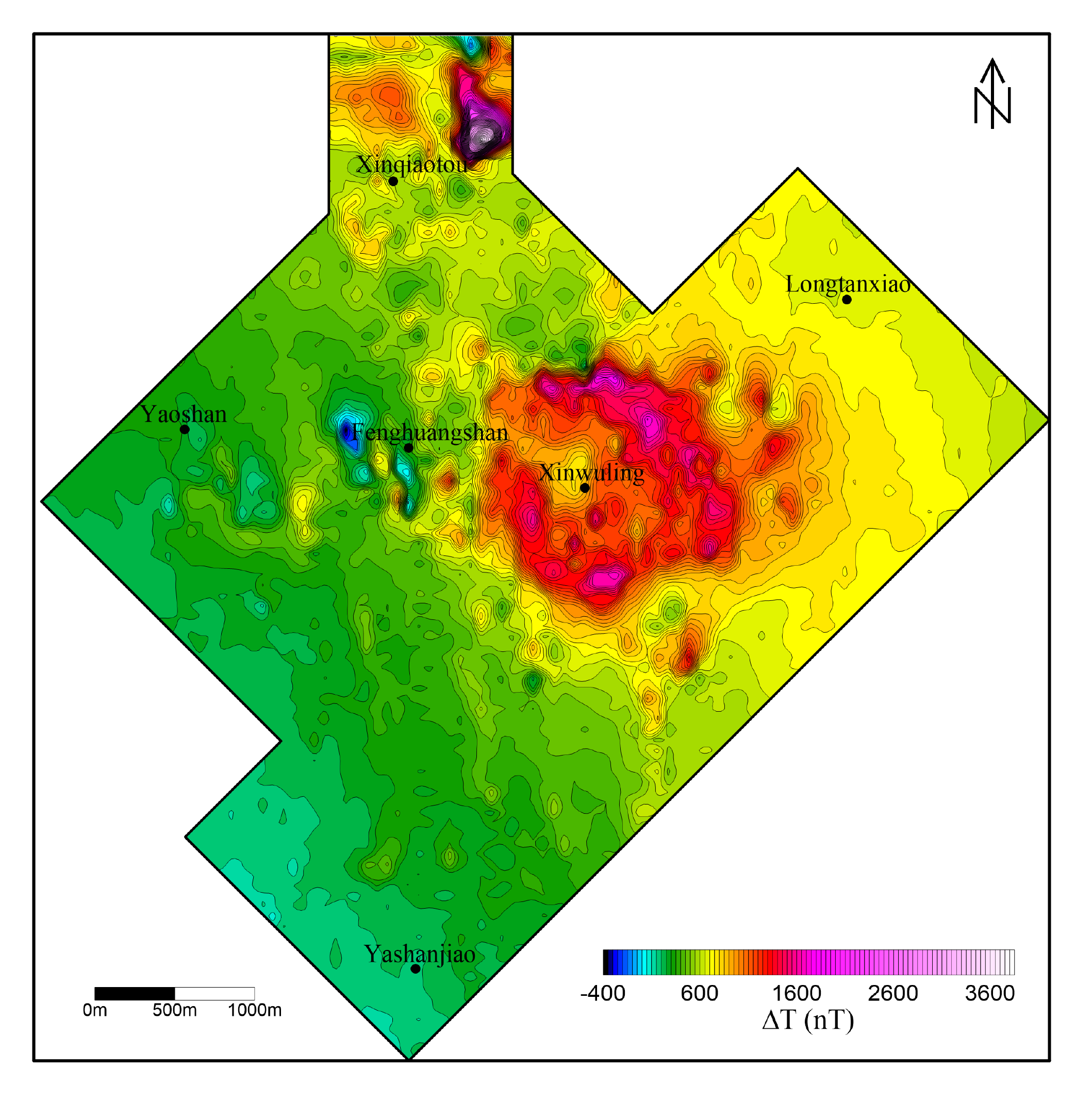}}
\caption{Figures~\ref{fig9a} and \ref{fig9b}  are the maps of the Bouguer gravity and the \texttt{RTP} magnetic anomalies of the study area in Tongling.} \label{fig9}
\end{figure}

\begin{figure}
\subfigure[]{\label{fig10a}\includegraphics[width=.49\textwidth]{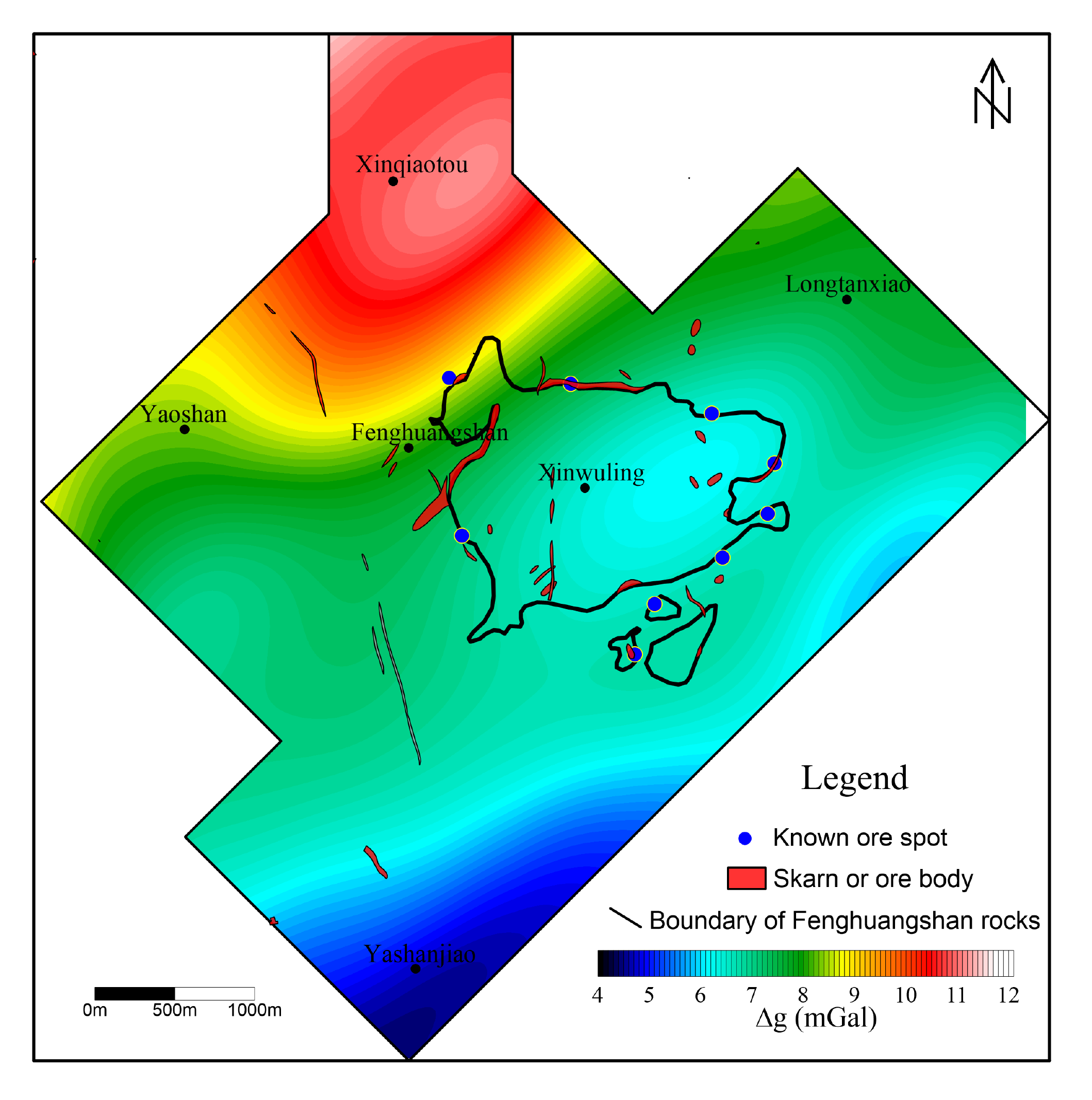}}
\subfigure[]{\label{fig10b}\includegraphics[width=.49\textwidth]{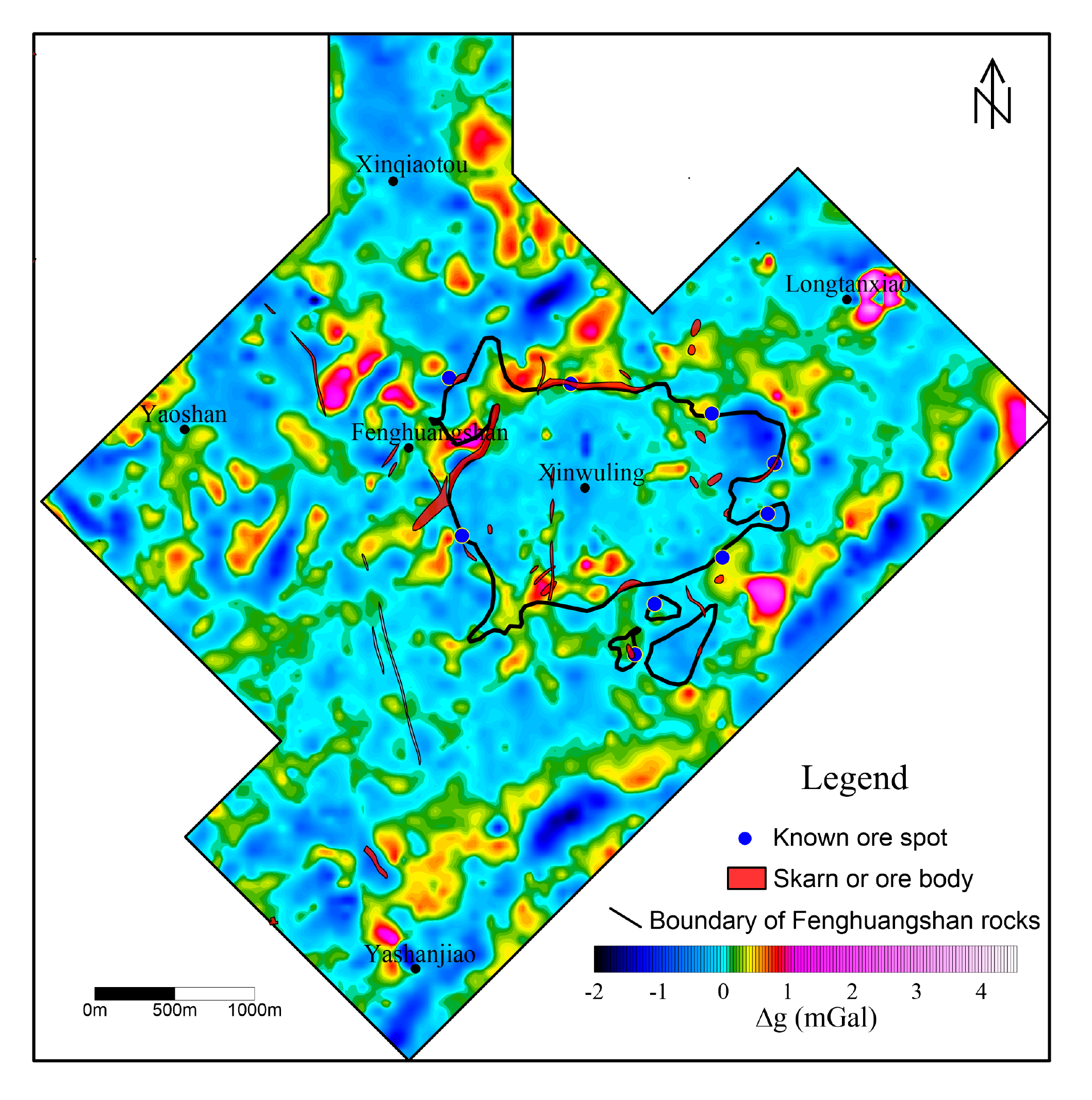}}\\
\subfigure[]{\label{fig10c}\includegraphics[width=.49\textwidth]{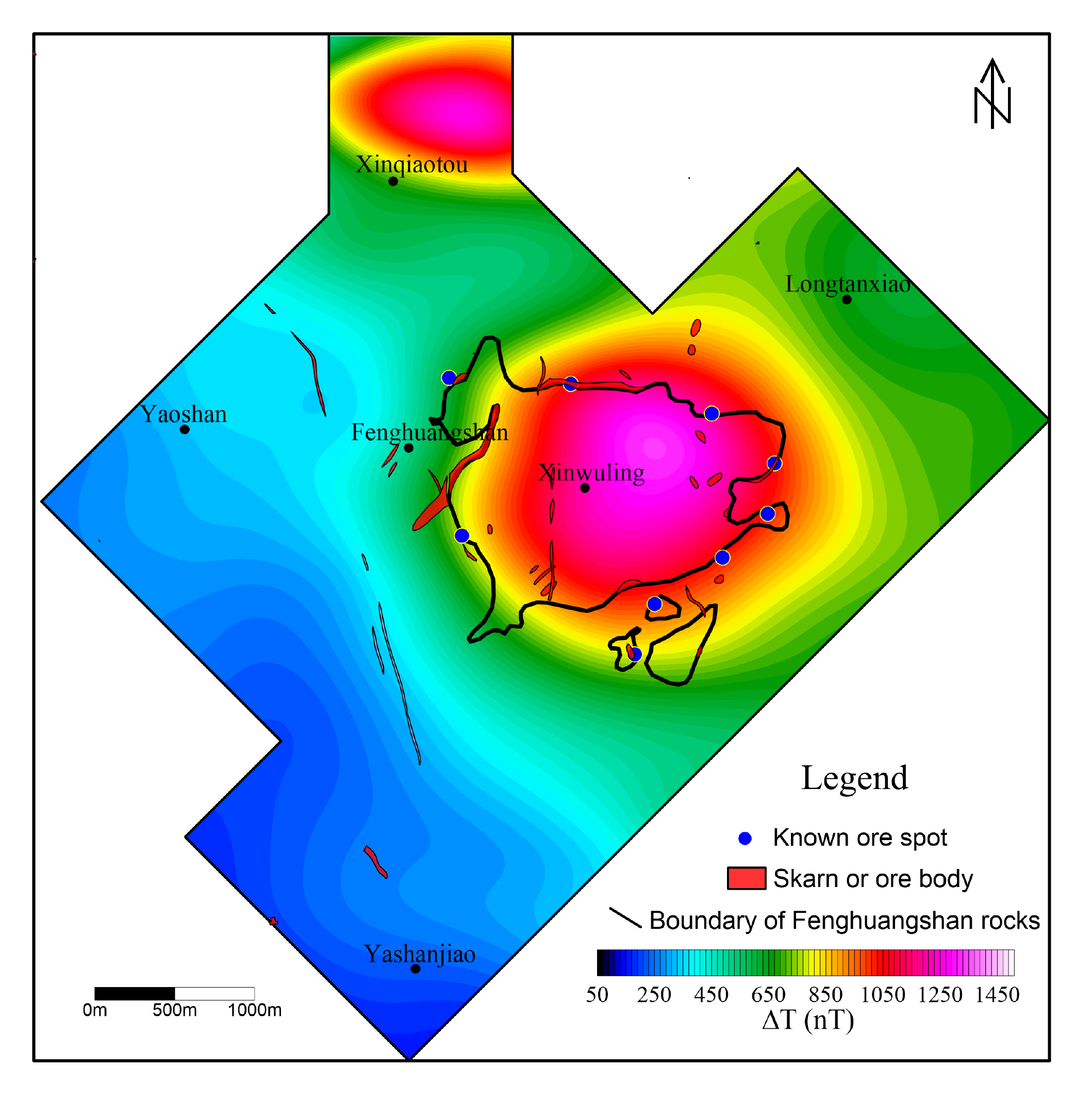}}
\subfigure[]{\label{fig10d}\includegraphics[width=.49\textwidth]{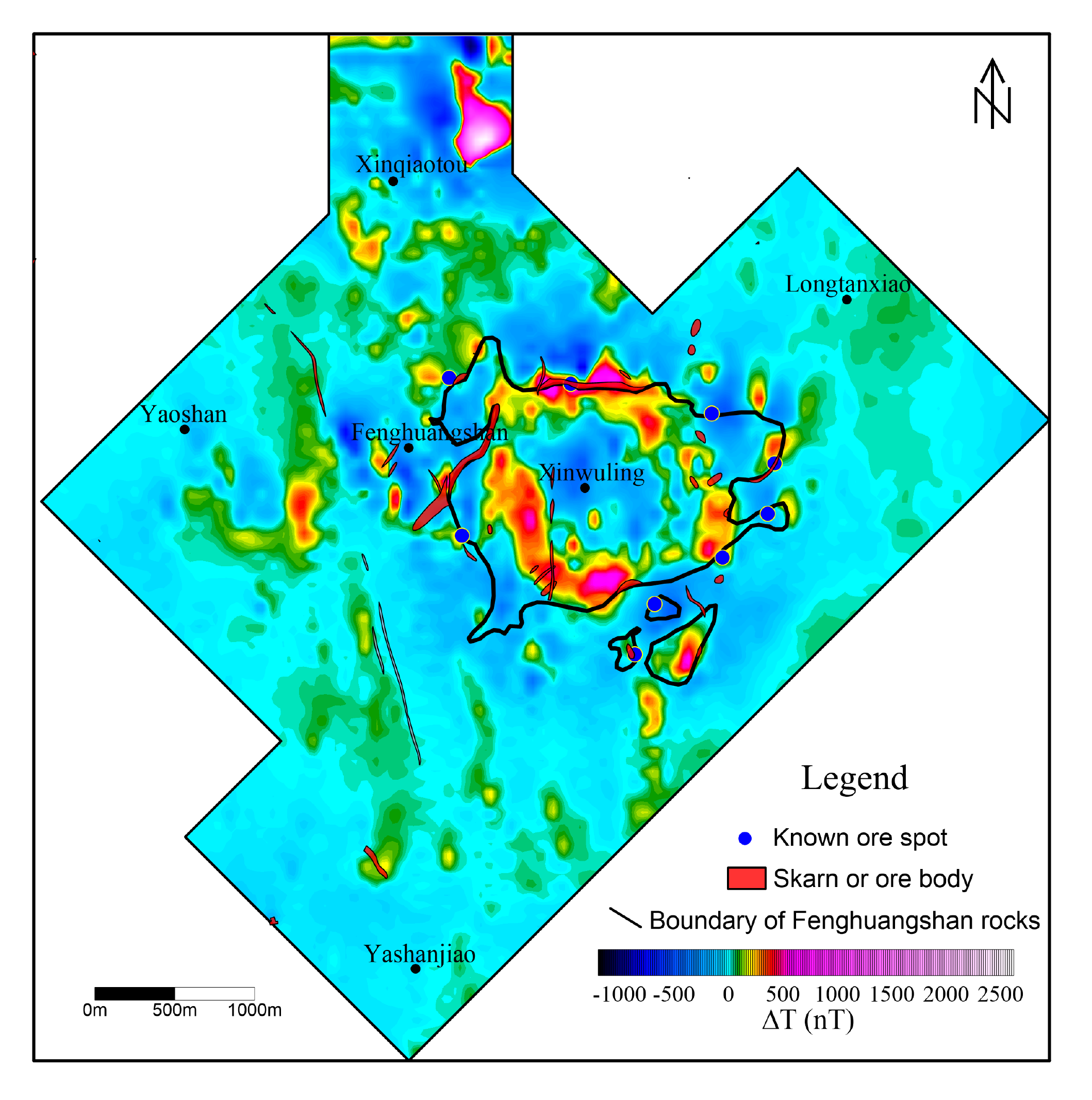}}
\caption{Figures~\ref{fig10a} and \ref{fig10b} are the separated regional and residual gravity anomalies of the study area, respectively; Figures~\ref{fig10c} and \ref{fig10d}  are the separated regional and residual magnetic anomalies of the study area, respectively.} \label{fig10}
\end{figure}

As we can see in Figure~\ref{fig10}, the separated regional gravity anomaly reflects the structures of deep underground sources, and it also reflects the distribution of the igneous rocks in the deep for the corresponding  local low gravity anomaly and known Fenghuangshan rocks. Due to the good correspondence of  the high magnetic anomaly with Fenghuangshan rocks, the separated regional magnetic anomaly mainly reflects the distribution of the igneous rocks in the deep. The gravity anomaly low and magnetic anomaly highs extend to the north-east of the Fenghuangshan rocks. Thus, the Fenghuangshan rocks in the deep are deduced to  extend to the north-east. The areas which correspond to local high gravity and magnetic anomalies in Figures~\ref{fig10b} and \ref{fig10d} are inferred to be skarns or shallow ore bodies which is consistent with known ore and skarn located in these areas. Therefore, we infer the unknown areas which may exhibit mineralizations based on the relations of the gravity and magnetic anomalies in Figures~\ref{fig10b} and \ref{fig10d}, as shown in Figure~\ref{fig11}.

\begin{figure}
\subfigure[]{\label{fig11}\includegraphics[width=.49\textwidth]{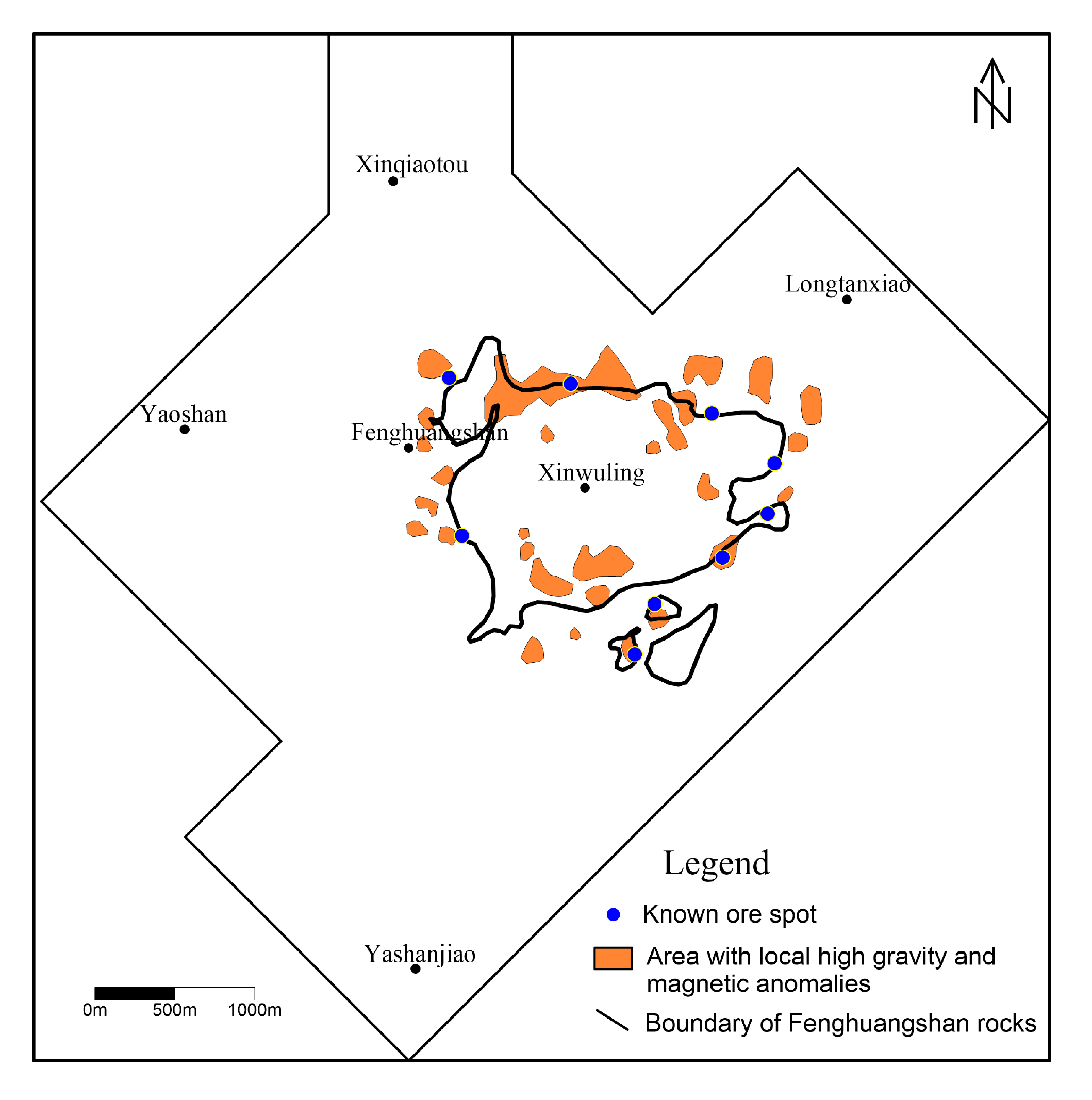}}
\caption{Predictions of the distributions of areas that may have sharns or ore bodies based on the separated high-gravity and high-magnetic fields.} \label{fig11}
\end{figure}


\section{Conclusions}\label{conclusion}
A fast non-convex low-rank matrix decomposition algorithm, \texttt{FNCLRMD\_PFS},  for the separation of potential field data has been presented and validated. The core of \texttt{FNCLRMD\_PFS}  is the efficient computation of the partial \texttt{SVD} of the block Hankel trajectory matrix, $\bfT$, without direct construction of $\bfT$. Thus, the low-rank matrix decomposition  non-convex algorithm for potential field data separation can be realized without requiring the storage of construction $\bfT$, and the resulting  storage and computational costs are lower than required when using  \texttt{LRMD\_PFS}. \texttt{FNCLRMD\_PFS}  depends on two parameters, these are the estimate $r^*$ of the rank of the regional anomaly matrix, and a threshold parameter $\beta$. 
Synthetic experiments were used to obtain recommendations for the settings of these parameters. These show that a suitable default interval for adjusting $\beta$ is $0< \beta <1/\sqrt{\mathtt{max}(KL,\hat{K}\hat{L})}$. The parameter $r^*$, when it is not too small,  mainly influences the computational time but not the accuracy. The experimental results demonstrate that the presented algorithm is robust and, thus, the choice of parameters, provided the interval for $\beta$ and rank $r^*$ are chosen as recommended, is straightforward.

Synthetic data sets were set up for gravity and magnetic data and used to contrast the accuracy and computational cost of \texttt{FNCLRMD\_PFS}  with \texttt{LRMD\_PFS}. These results demonstrated that \texttt{FNCLRMD\_PFS}  has higher accuracy and is more computationally efficient than \texttt{LRMD\_PFS}. Moreover, it is feasible to use \texttt{FNCLRMD\_PFS}  for matrices of  much larger size than is possible with \texttt{LRMD\_PFS} which exhibits either with an extreme requirement on  computational time or the report of  ``out of memory" for matrices of large size. Specifically,  \texttt{FNCLRMD\_PFS}  can be used to compute  large size potential field data with high accuracy at acceptable  computational cost. Finally,  \texttt{FNCLRMD\_PFS}  was also used for the separation of real data in the Tongling area, Anhui province, China. The separated low-gravity and high-magnetic regional anomalies have good correspondence  to the igneous rocks, and the separated high-gravity and high-magnetic residual anomalies exhibit  good correspondence to the known ore spots. Consequently, unknown areas of mineralizations can be inferred from the separated anomalies.

 


\bibliographystyle{plainnat}

\appendix
\section{Notation}\label{Notations}
Acronyms and notation used throughout are provided in Tables~\ref{Acronyms table} and \ref{Notations table}. 
\begin{table}

\captionof{table}{Acronyms used throughout \label{Acronyms table}}
\begin{tabular}{l l c |c}
Acronym& Description \\ \hline
\texttt{FBHMRSVD} & fast block Hankel matrix randomized SVD algorithm\\
\texttt{FBHMMM} & fast block Hankel matrix-matrix multiplication algorithm\\
\texttt{FBHMVM} & fast block Hankel matrix-vector multiplication Algorithm\\
\texttt{FNCLRMD\_PFS} & fast non-convex low-rank matrix decomposition algorithm for potential field separation\\
\texttt{EALM}  & exact augmented Lagrange multiplier method \\
\texttt{IALM}  & inexact augmented Lagrange multiplier method\\
\texttt{LRMD\_PFS} & low-rank matrix decomposition for  potential field separation\\
\texttt{RPCA}& robust principal component analysis\\
\texttt{RSVD} & randomized singular value decomposition\\
\texttt{SVD} & singular value decomposition\\
\texttt{RMSE} & root mean square error \\
\texttt{RTP} & reduce to the pole 
\end{tabular}

\end{table}
\begin{table}

\captionof{table}{Notation used throughout \label{Notations table}}
\begin{tabular}{l l c |c}
Notation& Description \\ \hline
$\bfJ$& exchange matrix  \\
$\bfX$& 2D gridded potential field data matrix\\
$\bfTj$& Hankel matrix constructed from the $j$th column of $\bfX$\\
$\bfT$& trajectory matrix of $\bfX$\\
$\bfX_1,\cdots,\bfX_Q$&  first to $Q$th columns of $\bfX$, respectively\\
$\bfU,\bfV,\bfSigma$& \texttt{SVD} of $\bfT$,  $\bfT=\bfU\bfSigma V^T$\\
$\bfU_r,\bfV_r,\bfSigma_r$&  rank-$r$ partial \texttt{SVD} of $\bfT$ using \texttt{FBHMRSVD}\\
$\bfXD$, $\bfXS$& data matrices of regional and residual anomalies, respectively\\
$\bfTD$, $\bfTS$&  trajectory matrices of $\bfXD$ and $\bfXS$, respectively\\
$\bfXD^*$, $\bfXS^*$& approximations of $\bfXD$ and $\bfXS$ using  \texttt{FNCLRMD\_PFS}, respectively\\
$\bfu_1,\bfu_2,\cdots$& $\bfU=[\bfu_1,\bfu_2,\cdots]$, $\bfu_1,\bfu_2,\cdots$ are the left singular vectors of $\bfT$\\
$\bfv_1,\bfv_2,\cdots$& $\bfV=[\bfv_1,\bfv_2,\cdots]$, $\bfv_1,\bfv_2,\cdots$ are the right singular vectors of $\bfT$\\
$x_{mn}$& element at $m$th row and $n$th column of $\bfX$\\
$P$, $Q$& $\bfX$ is of size $P \times Q$\\
$K$, $L$&  $\bfTj$ is of size $K \times L$\\
$\hat{K}$, $\hat{L}$& $\bfT$ is a block Hankel matrix with $\hat{K} \times \hat{L}$ blocks\\
$P_C$, $Q_C$& $\bfC$ is of size $P_C \times Q_C$\\
$\sigma_1$, $\sigma_2$, $\cdots$& $\bfSigma=\mathtt{diag}(\sigma_{1}^{2},\sigma_{2}^{2},\cdots)$, where $\sigma_1$, $\sigma_2$ , $\cdots$ are the singular values of $\bfT$\\
$r$& desired rank parameter in  \texttt{FBHMRSVD}\\
$p$& oversampling parameter in  \texttt{FBHMRSVD}\\
$q$& power iteration parameter in  \texttt{FBHMRSVD}\\
$r^*$& desired rank parameter in \texttt{FNCLRMD\_PFS}\\
$\beta$& thresholding parameter in \texttt{FNCLRMD\_PFS}\\
$\alpha$& weighting parameter in \texttt{LRMD\_PFS}\\
$\|\cdot\|_p$, $\|\cdot\|_*$& $\ell_p$ and nuclear norms, respectively\\
$T_{\mathtt{SVD}}$& computational cost of \texttt{SVD}\\
$T_{\mathtt{RSVD}}$& computational cost of \texttt{RSVD}\\
$T_{\mathtt{FBHMRSVD}}$& computational cost of \texttt{FBHMRSVD}\\
\end{tabular}

\end{table}

\section{The impact of the choice of the FFT used by MATLAB  on the computational cost}\label{FFT cost of MATLAB}
Our initial investigation of the computational cost of Algorithm~\ref{FBHMVMAlgorithm} demonstrated a general tendency for the computational cost to increase monotonically with increasing size of the matrices. There were, however, outlier sizes which were significantly higher in cost and departed from the general monotonic increase in time.  This is illustrated in Figure~\ref{figA1} for which  we conducted an experiment to test the cost of step~\ref{VMstep1} in Algorithm~\ref{FBHMVMAlgorithm} using the $\myvec(\bfX)$ with its dimensions between $2^{15}$ and $80000$.  For each matrix dimension, the code is run $80$ times, and the average time is calculated. But, because the MATLAB function determines an optimal transform to use for a given matrix size, at greater cost in the first run, this first run is excluded from the estimate of the average cost for each matrix size.   A spike in cost is seen between $60000$ and $70000$, actually at $63001$, but overall the tendency is a gradual increase in computational cost and outliers are not frequent.  We note that $63001=251 \times 251$ is not prime but $251$ is prime, and the determination of an optimal transform depends on the factorization of the transform size.  We conclude that there may be cases where the computational cost of Algorithm~\ref{FBHMVMAlgorithm} spikes because of this situation. On the other hand, for the problem of this size the calculation of the \texttt{RSVD} with, and without, the use of Algorithm~\ref{FBHMMMAlgorithm}  for the matrix multiplications has a computational cost in each case of  $2.614$ s and $30.885$ s, respectively. Hence, even when the FFT transform is relatively slow, the use of a fast block Hankel matrix multiplication is still faster than the use of a direct matrix-multiplication without the use of the FFT.

\begin{figure}
\subfigure{\label{figA1}\includegraphics[width=.65\textwidth]{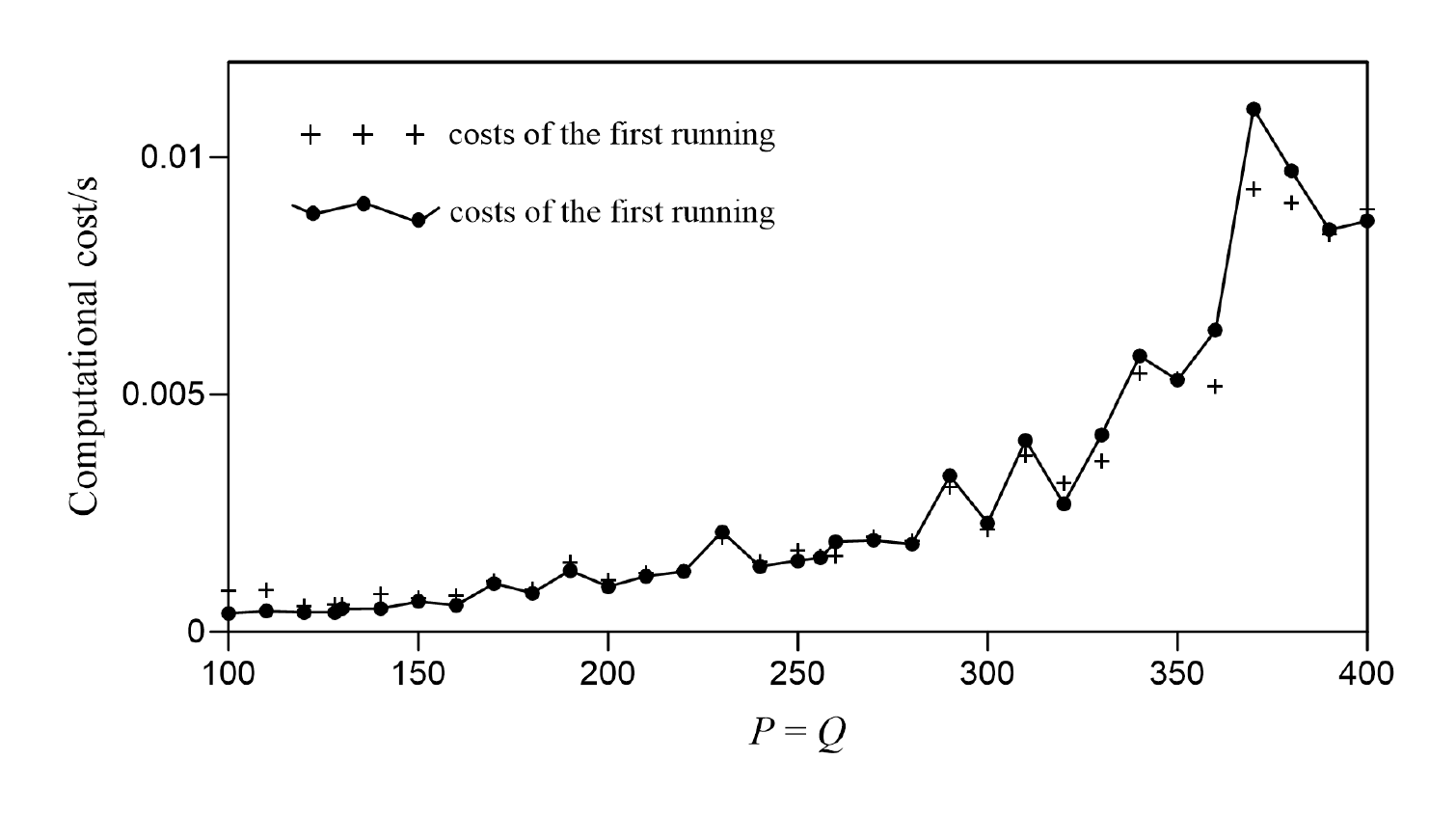}}
\caption{Demonstrating non-monotonic increase in computational time using  Algorithm~\ref{FBHMVMAlgorithm} for $P\times Q$ between $2^{15}$ and $80000$. \label{figA1} }
\end{figure}


\section*{Acknowledgments} Dan Zhu and Hongwei Li acknowledge the support of the National Key R\&D Program of China (2018YFC1503705). 
Rosemary Renaut acknowledges the support of NSF grant  DMS 1913136:   ``Approximate Singular Value Expansions and Solutions of
Ill-Posed Problems". 
Hongwei Li acknowledges the support of Hubei Subsurface Multi-scale Imaging Key Laboratory (China University of Geosciences) (SMIL-2018-06).
We also acknowledge Anhui Geology and Mineral Exploration Bureau 321 Geological Team for providing the data sets that were used for the real data experiments.

\medskip

\end{document}